\documentclass[prx,aps,twocolumn,superscriptaddress]{revtex4-2}
\usepackage{csquotes}
\MakeOuterQuote{"}

\usepackage{slashed}
\usepackage{float}
\usepackage[dvipsnames,table]{xcolor}

\usepackage{tikz-cd}

\makeatletter
\usepackage[caption=false]{subfig}

\usepackage[pdftex,colorlinks=true]{hyperref}
\hypersetup{linkcolor=blue,citecolor = blue,urlcolor=blue}
\usepackage{bbm}
\usepackage{graphicx}

\usepackage{mathtools}
\usepackage{stmaryrd}
\usepackage[force]{feynmp-auto}

\usepackage{bbold}	% for \mathbb{1} as a nice unit matrix symbol

\usepackage[normalem]{ulem}        % strike-through text
\usepackage{amsmath,mathtam} % mathtam includes math packages as well as nifty physics-related commands

\usepackage{booktabs} % To thicken table lines; LOAD BEFORE XCOLOR
\usepackage{setspace}
\usepackage{multirow} % merge cells vertically in tables
\usepackage[makeroom]{cancel} % crossed terms in tables

\definecolor{tbcolor}{HTML}{E1EFF0}

\newcolumntype{x}[1]{>{\centering\let\newline\\\arraybackslash\hspace{0pt}}p{#1}}

\makeatletter
\newcommand*\rel@kern[1]{\kern#1\dimexpr\macc@kerna}
\newcommand*\widebar[1]{%
  \begingroup
  \def\mathaccent##1##2{%
    \rel@kern{0.8}%
    \overline{\rel@kern{-0.8}\macc@nucleus\rel@kern{0.2}}%
    \rel@kern{-0.2}%
  }%
  \macc@depth\@ne
  \let\math@bgroup\@empty \let\math@egroup\macc@set@skewchar
  \mathsurround\z@ \frozen@everymath{\mathgroup\macc@group\relax}%
  \macc@set@skewchar\relax
  \let\mathaccentV\macc@nested@a
  \macc@nested@a\relax111{#1}%
  \endgroup
}
\newcommand\figref{Fig.~\ref}
\makeatother
	
\makeatletter

\newcommand*\wthelper[2]{%
    \hbox{\dimen@\accentfontxheight#1%
        \accentfontxheight#11.15\dimen@
        $\m@th#1\widetilde{#2}$%
        \accentfontxheight#1\dimen@
    }%
}
\newcommand*\accentfontxheight[1]{%
    \fontdimen5\ifx#1\displaystyle
        \textfont
    \else\ifx#1\textstyle
        \textfont
    \else\ifx#1\scriptstyle
        \scriptfont
    \else
        \scriptscriptfont
    \fi\fi\fi3
}
\makeatother

	\DeclareMathAlphabet{\mathbbold}{U}{bbold}{m}{n}
	\def\abs#1{\left|{#1}\right|}      	% absolute value, \abs{x} gives |x|
	\def\ket#1{\left|{#1}\right>}				% "ket"-state

	\def\braket#1#2{\left<{#1}|{#2}\right>}				% "bra-ket"-product
	\def\bs{{\bm{s}}}
	\def\bh{{\bm{h}}}

	\newcounter{subeqn} %
	\makeatletter
	\@addtoreset{subeqn}{equation}
	\makeatother

\definecolor{XQ}{rgb}{1,0.5,0}
\definecolor{XH}{rgb}{.5,0,0}
\definecolor{XL}{rgb}{0,0,1}

\begin{document}
\title{Entanglement Features of Random Neural Network Quantum States}
\date{\today}
\author{Xiao-Qi Sun}
\affiliation{Department of Physics and Institute for Condensed Matter Theory,
University of Illinois at Urbana-Champaign, Urbana, Illinois 61801, USA}
\author{Tamra Nebabu}
\affiliation{Stanford Institute for Theoretical Physics, Stanford University, Stanford, California 94305, USA}
\affiliation{Department of Applied Physics, Stanford University, Stanford, California 94305, USA}
\author{Xizhi Han}
\affiliation{Stanford Institute for Theoretical Physics, Stanford University, Stanford, California 94305, USA}
\affiliation{Kavli Institute of Theoretical Physics, University of California, Santa Barbara, California 93106, USA}
\author{Michael O. Flynn}
\affiliation{Department of Physics, Boston University, Boston, Massachusetts 02215, USA}
\author{Xiao-Liang Qi}
\affiliation{Stanford Institute for Theoretical Physics, Stanford University, Stanford, California 94305, USA}
\begin{abstract}
Restricted Boltzmann machines (RBMs) are a class of neural networks that have been successfully employed
as a variational ansatz for quantum many-body wave functions. Here, we develop an analytic method to study
quantum many-body spin states encoded by random RBMs with independent and identically distributed complex
Gaussian weights. By mapping the computation of ensemble-averaged quantities to statistical mechanics models,
we are able to investigate the parameter space of the RBM ensemble in the thermodynamic limit. We discover
qualitatively distinct wave functions by varying RBM parameters, which correspond to distinct phases in the
equivalent statistical mechanics model. Notably, there is a regime in which the typical RBM states have near-maximal
entanglement entropy in the thermodynamic limit, similar to that of Haar-random states. However,
these states generically exhibit nonergodic behavior in the Ising basis, and do not form quantum state designs,
making them distinguishable from Haar-random states.
\end{abstract}
\maketitle

\section{Introduction}
A fundamental challenge in the study of quantum many-body systems is the problem of efficiently describing physically interesting wavefunctions. In principle, a complete description of a quantum state requires classical resources that scale exponentially with system size. Fortunately in many cases, the set of physically relevant states occupy a small fraction of Hilbert space and can be parametrized efficiently. For instance, ground states of gapped local Hamiltonians, which typically exhibit area-law entanglement~\cite{masanes2009area, hastings2007area}, can be efficiently represented by matrix product states in one spatial dimension~\cite{Ostlund:1995,Rommer:1997} or by tensor network states~\cite{Verstraete:2008} in higher dimensions. However, it remains a daunting task to construct a suitable ansatz for more general many-body wavefunctions while efficiently capturing their salient macroscopic properties.

An analogous challenge is encountered in the field of machine learning, where one is tasked with learning classical probability distributions over high-dimensional spaces. 
The shared "curse of dimensionality," which besets both machine learning and quantum many-body physics, offers a tantalizing hint that similar techniques can be employed in both fields and has led to a growing interest in the application of machine learning techniques to physics (for review see \cite{Carleo:2019,Torlai:2020,Carrasquilla:2020} and references therein). In machine learning applications, neural networks serve as function approximators capable of extracting important correlations in high-dimensional spaces. Notably, there exist universal approximation theorems~\cite{Hornik:1989, zhou2020universality,hornik1993some,sonoda2017neural} that prove the expressive power of neural networks to approximate arbitrary multivariable classical functions.

Motivated by the success of neural networks in learning classical probability distributions over high-dimensional spaces, neural network ansatzes~\cite{Carleo:2017,Gao:2017,Carleo:2018,Saito:2018,Choo:2018,He:2019,Choo:2019,melko2019restricted,hartmann2019neural,Irikura:2020,Schmitt:2020,Adams:2021,Liang:2021} have been proposed to efficiently parameterize certain classes of many-body quantum states. The breadth of neural network architectures and parameters, and the inherent nonlinearity built into their functional form, allow them to serve as a more general variational ansatz than typical tensor networks \cite{chen2018equivalence}. Recent works~\cite{Carleo:2017,Carleo:2018,melko2019restricted} have shown that a particular category of elementary neural networks, known as restricted Boltzmann machines (RBMs), are capable of efficiently representing a large class of quantum states, including examples with maximal bipartite entanglement~\cite{Deng:2017,Levine:2019}.

It is therefore desirable to characterize the set of many-body wavefunctions with efficient RBM representations. Other ensembles of quantum states have been studied with analytical methods, such as Haar-random states~\cite{Page:1993} and random tensor network states~\cite{Hayden:2016}. Advances in the study of random circuits has also expanded the understanding of entanglement properties to states beyond traditional ansatz~\cite{Skinner:2019,2019MeasurementEntanglement,2020MeasurementCriticality,Bao:2020,2021EntanglementNegativity,2021EntanglementDomains}, and constrained systems have been shown to exhibit universal entanglement properties of their own~\cite{2020ConstrainedEntanglement}. In contrast, while entanglement transitions have been explored in a certain type of RBM ansatz~\cite{2021PhRvB.104j4205M}, the study of entanglement features of generic random RBM states has been limited to finite-size numerics~\cite{Deng:2017}.

In this paper, we analytically characterize many-body quantum spin states encoded by single-layer random Gaussian RBMs, and support our conclusions with finite-size numerics. Using techniques similar to those used in the study of random tensor networks \cite{Hayden:2016}, we are able to describe ensemble-averaged properties of the RBM states, such as their second R\'{e}nyi entropy and statistical fluctuations in the norm of the variational wavefunction. In particular, we find that the computation of such quantities may be mapped to classical statistical mechanical models of (coupled) spin chains of length $N$. Interestingly, in the thermodynamic limit $N \to \infty$, these models exhibit phase transitions as one varies the ensemble parameters. These transitions separate wavefunctions with qualitatively different properties. Using both analytics and numerics, we show that random Gaussian RBM states are able to exhibit maximal entanglement at leading order in $1/N$ for a particular regime of RBM parameters.  However, even in this special regime, the Gaussian RBM ensemble can be distinguished from the ensemble of Haar-random states and it does not form a quantum state design. %, and our numerical studies show that the entanglement spectrum features no level repulsion.}

The paper is organized as follows: In Sec.~\ref{sec:setup}, we introduce the construction of an RBM and its mapping to a quantum many-body state of $N$ spins, and establish the parameters of the RBM ensemble under consideration. In Sec.~\ref{sec:norm_fluc}, we study the norm fluctuations of the RBM ensemble as a function of its parameters, and establish two regimes of qualitatively different behaviors, corresponding to vanishing and non-vanishing norm fluctuations to lowest order in $1/N$. In Sec.~\ref{sec:S2}, we study the ensemble-averaged second R\'{e}nyi entanglement entropy and construct a phase diagram over the parameter space of the RBM ensemble. We report our findings of a near-maximal entanglement regime in the large-$N$ limit, and support these analytic results with finite-size numerics. In Sec.~\ref{sec: other_measure}, we investigate other properties of the RBM states such as their von Neumann entanglement entropy, entanglement spectrum, fractal dimensions, and their relation to quantum state designs. Importantly, we show that the fractal dimensions of the wavefunctions in the Ising basis can distinguish the near-maximal RBM states from generic random states and explain the entanglement suppression for large number of hidden neurons. %\TN{Mention content of new subsection?} 
We summarize our findings and carry out further discussions in Sec.~\ref{sec:conclusions}.

\section{Construction of the RBM ensemble}
\label{sec:setup}
\begin{figure}[t]
    \centering
    \includegraphics[width=0.8\linewidth]{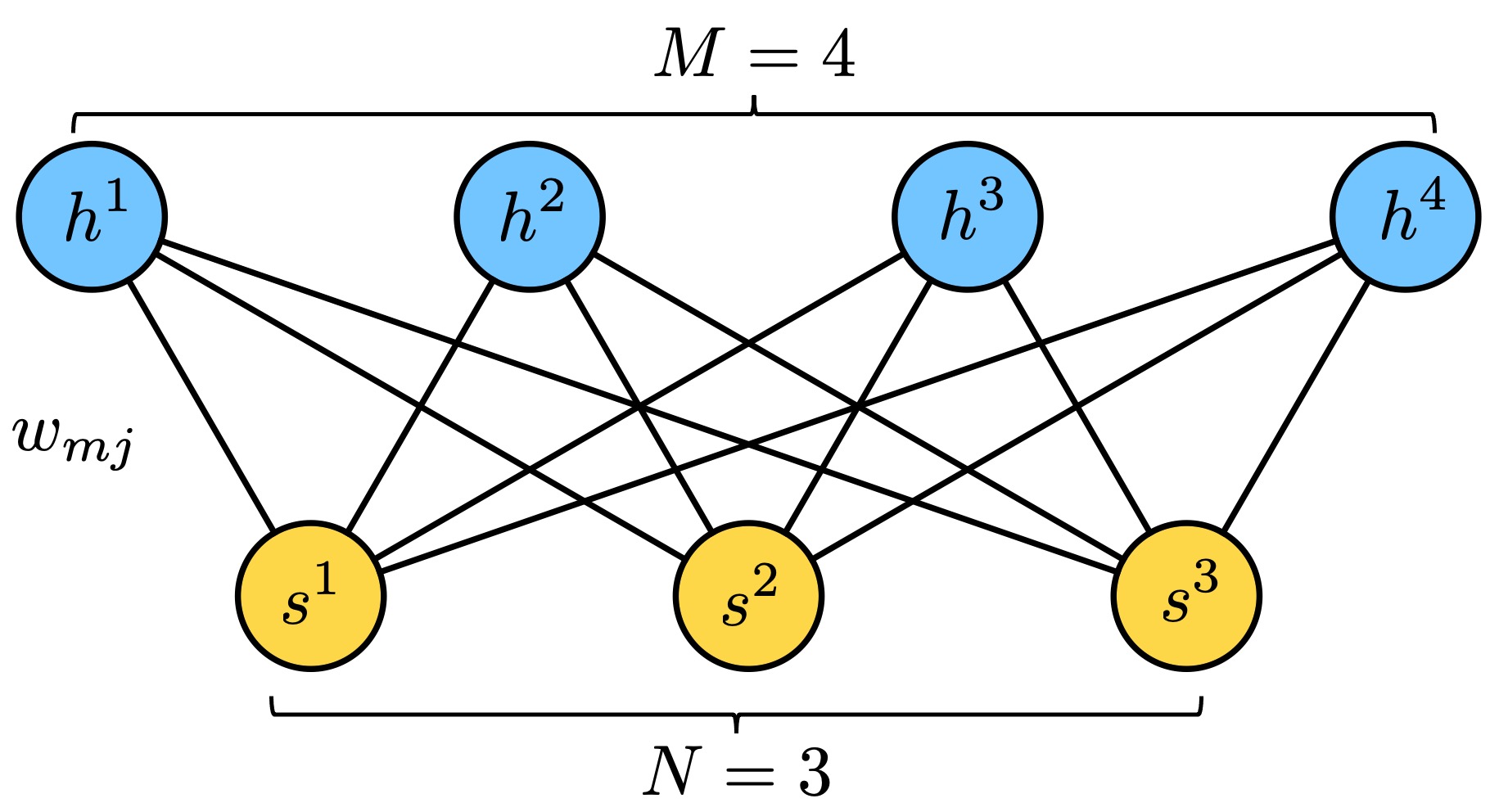}
    \caption{A schematic of the RBM architecture considered: There are $N$ visible neurons in the first layer labeled by $s^j$ and $M$ hidden neurons in the second layer labeled by $h^m$. The two layers are all-to-all connected with weights $w_{m j}$. This architecture will be used to encode wavefunctions of $N$ physical quantum spins.}
    \label{fig: RBM_wf}
\end{figure}

We begin by introducing the mapping between an RBM configuration and a quantum spin state. An RBM is a two-layer, bipartite neural network model where the visible neurons are located in the first layer and the hidden neurons are located in the second layer (see Fig.~\ref{fig: RBM_wf}). Edges, which generate interactions between the neurons, are permitted \textit{between} the two layers, but not within the layers themselves. We will choose the neurons to be binary variables taking on the values $\{-1,1\}$. There will be $N$ visible neurons, corresponding to physical spin variables, and $M$ hidden neurons, corresponding to auxiliary spin variables.
The parameters of an RBM consist of a collection of weights $w_{mj}$ that represent interactions between the visible and hidden neurons, as well as onsite potentials $a_j$, $b_m$ on the visible and hidden neurons, respectively. The parameters $\SW = \{w_{mj},a_j,b_m\}$ are generically complex. The configuration of an RBM is then mapped to an \textit{unnormalized} many-body quantum state $\ket{\Psi}$ of $N$ spins as follows:
\begin{equation}
    \ket{\Psi} \equiv    \frac{1}{2^M} \sum_{\bh,\bs} e^{-\sum_j a_j s^j - \sum_m b_m h^m - \sum_{jm} w_{mj} s^j h^m }\ket{\bs},
    \label{eq: RBM_wf}
\end{equation}
where $\bs$ and $\bh$ are shorthands for the many-body configurations of visible layer spins $\bs \equiv \{s^1,s^2,...,s^N\}$ and hidden layer spins $\bh\equiv\{h^1,h^2,...,h^M\}$. We will reserve $m$ (and $m'$) as indices for the hidden layer neurons in order to distinguish them from the physical spin indices. 

For simplicity, we will restrict our attention to the zero-bias case ($a_j = b_m = 0$). However, the analytic method that we shall develop is applicable more generally. In the zero-bias case, the simplifed expression for the unnormalized wavefunction encoded by an RBM is
\begin{equation}
    \ket{\Psi} =\sum_{\bs} \prod_{m=1}^M \cosh\left( \sum_{j}w_{mj} s^j\right)\ket{\bs},%\equiv \sum_{\bs}\Psi(w,\bs)|\bs\rangle,
  \label{eq: RBM_wf_zerobias}
\end{equation}
where we have explicitly summed over the hidden spin indices.

We shall consider an ensemble of RBMs obtained by drawing the network weights from a probability distribution. In paritcular, we will take the weights to be independent and identically distributed (iid) random complex variables.
Because the RBM wavefunction is unnormalized by convention, the average norm and the norm fluctuations will also be properties of the ensemble. We can write the random weight as $w_{mj} = w_{mj}^R +i w_{mj}^I$ where the real part and imaginary part are independently drawn from Gaussian distributions of mean 0 and variances $\sigma_R^2$ and $\sigma_I^2$, respectively. 

We now define an appropriate thermodynamic (large-$N$) limit of the RBM wavefunction by considering the $N$-dependent scaling of the available parameters. For a fixed spin configuration, the wavefunction amplitude is the product of $M$ independent random variables $\cosh(\sum_{j}w_{m j}s^{j})$ introduced by $M$ hidden spins. In order to define an appropriate large-$N$ limit, the scaling of the variances should be chosen such that $\sum_{j}w_{m j}s^{j}$ remains an $\OO(1)$ number as $N$ increases. This implies $\sigma_R^2=u^2/N$ and $\sigma_I^2=v^2/N$ for some set of $\mathcal{O}(1)$ constants $u$ and $v$. As such, $u$, $v$, and $\lambda \equiv M/N$ (the ratio of hidden/auxiliary spins to visible/physical spins) will span the three-dimensional parameter space of our RBM ensemble.

Having finished the construction of our RBM ensemble, let us highlight some of its important properties. One feature that will have important consequences later is the qualitatively different roles of $u$ and $v$: as one can see in Eq.~\eqref{eq: RBM_wf_zerobias}, the imaginary part of the weights appearing in the argument of $\cosh()$ does not modulate the norm of the wavefunction (or its fluctuations) as strongly as the real part of the weights. This will ultimately lead us to investigate the limit $(u,v)=(0,\infty)$, which has the intriguing property that the half-system entropy of the RBM wavefunction approaches the maximal value of $(N/2) \log 2$~\cite{Page:1993} to leading order in large $N$ for a certain range of $\lambda$. Another important feature is the symmetry of the RBM ensemble. The components of the weights $w_{m j}^{R,I}$ are iid random variables drawn from a real, symmetric distribution; hence the ensemble has a $\mathbb{Z}_2$ symmetry for each individual weight and also a permutation symmetry between all weights.

Lastly, we remark on a general property of \emph{all} RBM states which will be relevant to the entanglement characteristics we will study later: namely, the entanglement entropy of a subsystem is upper bounded by $M\log 2$ when $M$ is smaller than the subsystem size. This is because the RBM wavefunction may be written as a superposition of at most $2^M$ product states, as seen by summing over the physical spin states in Eq.~\eqref{eq: RBM_wf}:
\begin{equation}
   \ket{\Psi} =\frac{1}{2^M}\sum_{\bh}e^{-\sum_m b_m h_m} \left(\ket{\psi_j^\bh}\right)^{\otimes_j},
   \label{eq: wf_product_states}
\end{equation}
where
\begin{equation}
    \ket{\psi_{j}^{\bh}}\equiv e^{-a_j-\sum_m h_m w_{mj}}\ket{\uparrow}_j+e^{a_j+\sum_m h_m w_{mj}}\ket{\downarrow}_j
\end{equation}
is a single spin state at site $j$ that depends on the hidden neuron states. Since $\left(\ket{\psi_j^\bh}\right)^{\otimes_j}$ is a direct product state, we conclude that the Schmit decomposition for any bipartition of the system contains at most $2^M$ terms. In other words, the rank of the reduced density operator for all subsystems are bounded from above by $2^M$. Consequently, the von Neumann entropy and all R\'enyi entropies are upper bounded by $M\log 2$.

\section{Norm fluctuation and phase transitions}
\label{sec:norm_fluc}
Since the wavefunction encoded by the Gaussian RBM ensemble is unnormalized, we begin by discussing the computation of the norm and its fluctuations. Because the square of the norm is a slightly easier quantity to work with, we compute $\overline{Z_0}\equiv\overline{\braket{\Psi}{\Psi}^2}$ and compare it to $\Big(\overline{\braket{\Psi}{\Psi}}\Big)^2$. The latter may be computed exactly from the average of the squared norm:
\begin{equation}
\begin{split}
    \overline{\braket{\Psi}{\Psi}}&=\sum_\bs\prod_{m}\overline{\cosh(W^{m}_{ \bs})^{*}\cosh(W^{m}_ \bs)}\\
    &=\frac{2^N}{2^M}\left[\exp(2u^2)+\exp(-2v^2)\right]^M,
\end{split}
\end{equation}
where, for convenience, we have defined the random variable $W^{m}_{ \bs}=\sum_{j}w_{m j}s^{j}$. The average was performed by taking advantage of the independence of the random weights and then using Gaussian integrals (see Appendix~\ref{app:norm_square_av}). For later convenience, it is useful to recast the above formula as
\begin{equation} 
\begin{split}\label{eq: logAvgNormSq_red}
\frac{\log\overline{\braket{\Psi}{\Psi}}}{N}=\lambda(u^2-v^2)&+\lambda \log\cosh(u^2+v^2)+\log 2.
\end{split}
\end{equation}
Meanwhile, we can compute $\overline{Z_0}\equiv\overline{\braket{\Psi}{\Psi}^2}$ by explicitly substituting the expression for the wavefunction components in Eq.~\eqref{eq: RBM_wf},
\begin{equation}
    \overline{Z_0}=
    \sum_{\bs_1,\bs_2}\prod_{m}\overline{\cosh(W^{m}_{\bs_1})^{*}\cosh(W^{m}_{ \bs_2})^{*}\cosh(W^{m}_{\bs_1})\cosh(W^{m}_{\bs_2})},
\end{equation}  
where $\bs_1$ and $\bs_2$ are two spin configurations. The ensemble average over Gaussian weights may be computed analytically as shown in Appendix~\ref{app: Z0}. Here, we simply present the result of the calculation:
\begin{equation}
\begin{split}
    \overline{Z_0} &=e^{2M(u^2-v^2)}\sum_{\bs_1,\bs_2}\Big{[}\frac{1}{4}(2+e^{2(u^2+v^2)}\cosh 4\phi u^2\\
    &+e^{-2(u^2+v^2)}\cosh 4\phi v^2) \Big{]}^M.
\end{split}
\label{eq: Z0_Gaussian}
\end{equation}
As a shorthand, we have defined $\phi\equiv (\bs_1\cdot \bs_2)/N$, where $\bs_1\cdot \bs_2$ stands for $\sum_{i=1}^{N}s_1^{i}s_2^{i}$. The fact that $\overline{Z_0}$ depends on the spin configurations $\bs_1$ and $\bs_2$ only through their dot product $\phi$ comes from the symmetry of the ensemble; simultaneously permuting or flipping the spin variables in the two copies leaves $\overline{Z_0}$ unchanged. Ultimately, this is due to the aforementioned symmetries of the random weights; they are iid random variables, and their probability distribution possesses a $\mathbb{Z}_2$ symmetry $w_{mj} \overset{\text{iid}}{\sim} -w_{m j}$. We can further recast $\overline{Z_0}$ as a partition function of a statistical mechanics problem:
\begin{equation}
    \overline{Z_0}=\sum_{\phi} \exp[-N \mathcal{F}(\phi)],
    \label{eq: Z0_free_energy}
\end{equation}
where the sum is over all possible values of $\phi$, and the free energy density $\mathcal{F}(\phi)\equiv\mathcal{E}(\phi)-\mathcal{S}(\phi)$ contains two contributions: an "energetic" part $\mathcal{E}(\phi)$ from the interactions of the coupled spin configurations $\bs_1$ and $\bs_2$, and an "entropic" part $\mathcal{S}(\phi)$, which comes from the degeneracy of $(\bs_1,\bs_2)$ configurations that map to the same $\phi$. The energy density is given by
\begin{equation}
    \mathcal{E}(\phi)=-\lambda\left[\log \Omega(\phi)+2(u^2-v^2)-\log 2\right],
    \label{eq: Z0_energy}
\end{equation}
with
\begin{equation}
    \Omega(\phi)\equiv 1+\frac{e^{2(u^2+v^2)}}{2}\cosh 4\phi u^2+\frac{e^{-2(u^2+v^2)}}{2}\cosh 4\phi v^2.
\label{eq: omega_def}
\end{equation}
Meanwhile, the entropy density is given by
\begin{equation}\label{eq: Z0_entropy}
    \mathcal{S}(\phi)=-\frac{1-\phi}{2}\log \frac{1-\phi}{2}-\frac{1+\phi}{2}\log \frac{1+\phi}{2}+\log 2.
\end{equation}
where we have neglected terms of higher order in $1/N$.

The dominant contributions to $\overline{Z_0}$ will come from $\phi$ values which minimize the free energy $\SF(\phi)$; this in turn corresponds to finding $\phi$ such that $\CE(\phi)$ is small and $\CS(\phi)$ is large. Since the parameters of the RBM ensemble $(u,v,\lambda)$ also enter into the expression, one may anticipate that tuning such parameters can result in phase transitions associated to changes in the free energy minima. Importantly, the free energy density is symmetric under $\phi\rightarrow-\phi$, which will lead to classification of phases by this $\Z_2$ symmetry. It is worth commenting that phase transitions in RBM-encoded functions have been observed in other contexts, such as for classical probability distributions for machine learning tasks~\cite{Barra:2017}.

One can check that the entropy density $\CS(\phi)$ is maximized at $\phi=0$ while the energy density $\CE(\phi)$ is minimized at $\phi=\pm 1$. Given that the energy contribution is proportional to $\lambda$ and the entropy part is independent of $\lambda$, one expects that for certain fixed $(u,v)$, tuning $\lambda$ can result in a second-order phase transition which separates the entropy-dominated phase with a single minimum from the energy-dominated phase with two minima. These two phases are classified by symmetry: the energy density dominates in the $\mathbb{Z}_2$-symmetry-broken phase, while the entropy density dominates in the $\mathbb{Z}_2$-symmetric phase.

\begin{figure}[hbtp]
    \centering
    \includegraphics[width=\linewidth]{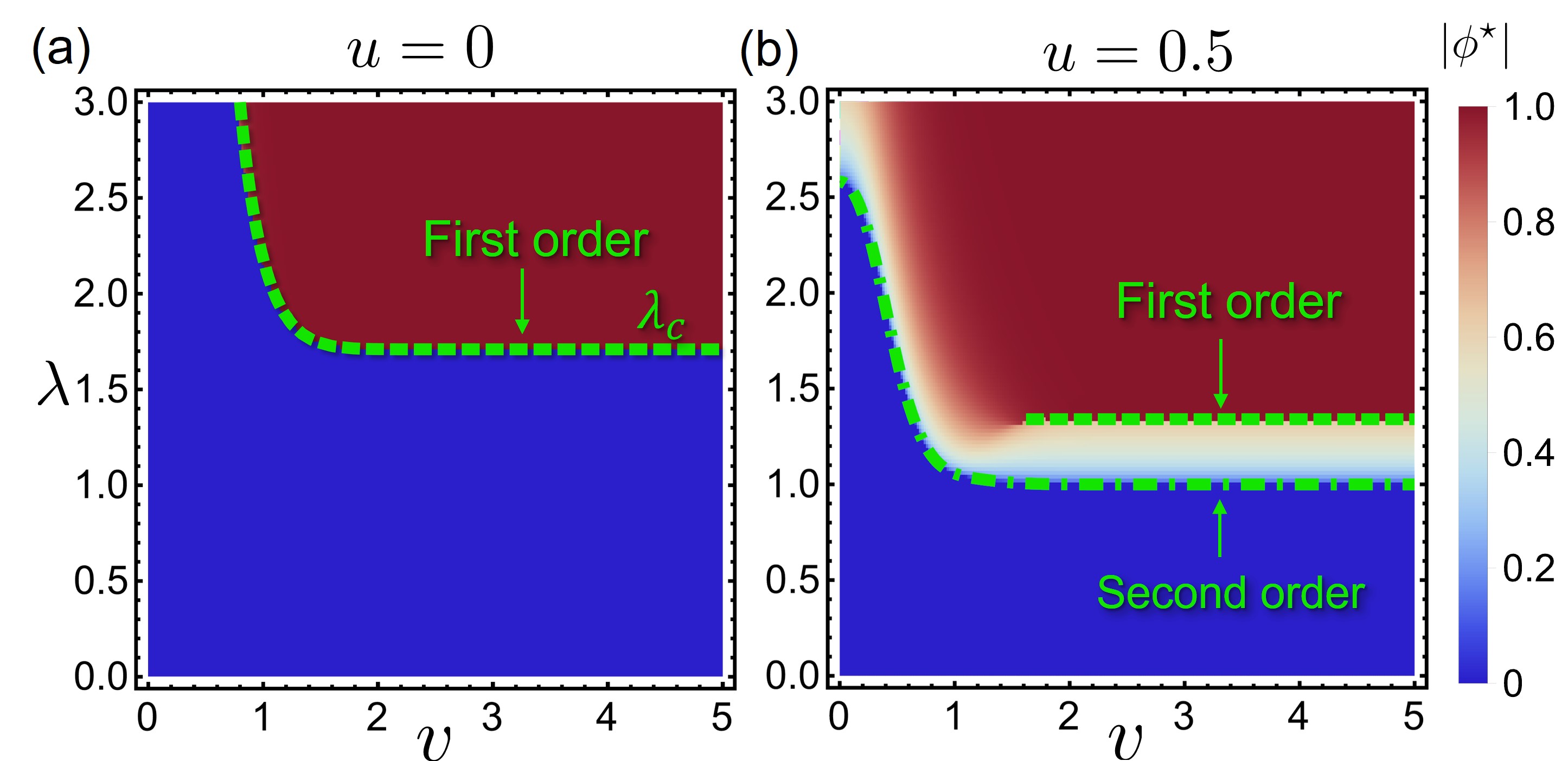}
    \caption{Phase diagrams for the statistical mechanics model associated to the calculation of $\overline{Z_0}$ for (a) $u=0$, and  (b) $u=0.5$. $\phi^{\star}$ represents the value of $\phi$ at a global minimum of the free energy.  In both cases, there are two phases which are classified by the $\mathbb{Z}_2$ symmetry of $\phi\rightarrow-\phi$: a phase with two free energy minima at $\phi^{\star} \simeq \pm 1$ (red) and another phase with $\phi^{\star} = 0$ (blue). For $u = 0$ in panel (a), the first-order phase boundary (green dashed line) at large $v$ approaches $\lambda=\lambda_c$, where $\lambda_c\equiv\frac{\log 2}{\log 3-\log 2}\approx 1.71$. In panel (b), both sides of the first-order transition spontaneously break the $\Z_2$ symmetry with nonzero $\phi^{\star}$, so that the first-order transition line can end at a second-order critical point. There is also a second-order transition (dashed-dotted line) at smaller $\lambda$.}
    \label{fig: phase_diagram_1}
\end{figure}

Furthermore, one can also deduce the existence of an additional first-order transition from the dependence of the energetic part on $(u,v,\lambda)$. For sufficiently large $v$ with $u<v$, the energy profile near $\phi=1$ possesses a sharp dip due to the second exponential term in Eq.~\eqref{eq: omega_def}. This gives rise to additional free energy minima that can become dominant over the local minimum at $\phi = 0$. Hence, we find an additional first-order phase transition when the new free energy minima formed at $|\phi|\simeq 1$ become the global minima. In Fig.~\ref{fig: phase_diagram_1}, we show cross sections of the phase diagram in the $\lambda-v$ plane for fixed $u=0$ and $u=0.5$. From the phase diagrams, one observes that the phase boundaries approach a constant $\lambda$ line as $v \to \infty$. (Notice that we take the large-$N$ limit before $v\to \infty$. )  

We will focus our attention on the limiting case of $u = 0$ and $v \to \infty$, which will be of interest later since it has a maximal entropy regime in the large-$N$ limit. For these parameters, we can explicitly work out the critical $\lambda_c$ associated with the phase boundary. We begin by writing a limiting expression for the energy density:
 \begin{equation}
    \mathcal{E}(\phi)=\begin{cases}
    \lambda\log 4, & \text{If } |\phi|<1;\\
    \lambda\log \frac 83, & \text{If } |\phi|=1,
    \end{cases}
\label{eq: energy_cases}
\end{equation}
where in each case, we have taken the dominating exponential factor in $\Omega(\phi)$. For a large but finite $v$, the minimum energy at $|\phi|=1$ increases quickly through a transition region of width $\sim 1/v^2$ and approaches its maximum value for $|\phi|=0$. The above expression approximates the sharply sloped energy near $|\phi|=1$ as a discontinuous function in the large-$v$ limit. As shown in Fig.~\ref{fig: phase_diagram_1}, the approximation already works sufficiently well for $(u,v)=(0,4)$ since the global minimum either is very close to $\pm 1$ (red regime) 
or is located at $0$ (blue regime).

We can compare the local minima of the free energy in the two cases. For $|\phi|<1$, the free energy is locally minimized at $\phi=0$ with $\mathcal{F}(\phi=0)=2\lambda\log 2-2\log 2$. On the other hand, for $|\phi|=1$, the free energy is $\mathcal{F}(\phi=\pm 1)=\lambda\log \frac 83-\log 2$. These two values for the minimum free energy coincide at the the phase boundary at $\lambda=\lambda_c\equiv \frac{\log 2}{\log 3-\log 2}$. At this critical $\lambda$, a first-order (symmetry-breaking) phase transition separates two phases.

Now we can compare the squared norm fluctuations of the two phases. For the phase dominated by the $\phi=0$ free energy minimum, we have
\begin{equation}
\begin{split}
\frac{\log\overline{Z_0}}{N}&\approx 2\lambda\left[\log \cosh(u^2+v^2)+(u^2-v^2)\right]\\
&+2\log 2.
\end{split}
\label{eq: Z0}
\end{equation}
Comparing Eq.~\eqref{eq: Z0} to Eq.~\eqref{eq: logAvgNormSq_red}, we can see that the squared norm fluctuations vanish to leading order in the large-$N$ limit, i.e., up to higher order terms in $1/N$, $\frac{1}{N}\log\overline{Z_0} \approx \frac{1}{N}\log \big(\overline{\braket{\Psi}{\Psi}}\big)^2$. This interesting result is specific to the $\mathbb{Z}_2$ symmetric phase. By contrast, in the symmetry breaking phase, the squared norm fluctuations are generically large. 

\begin{figure}[H]
    \centering
    \includegraphics[width=\linewidth]{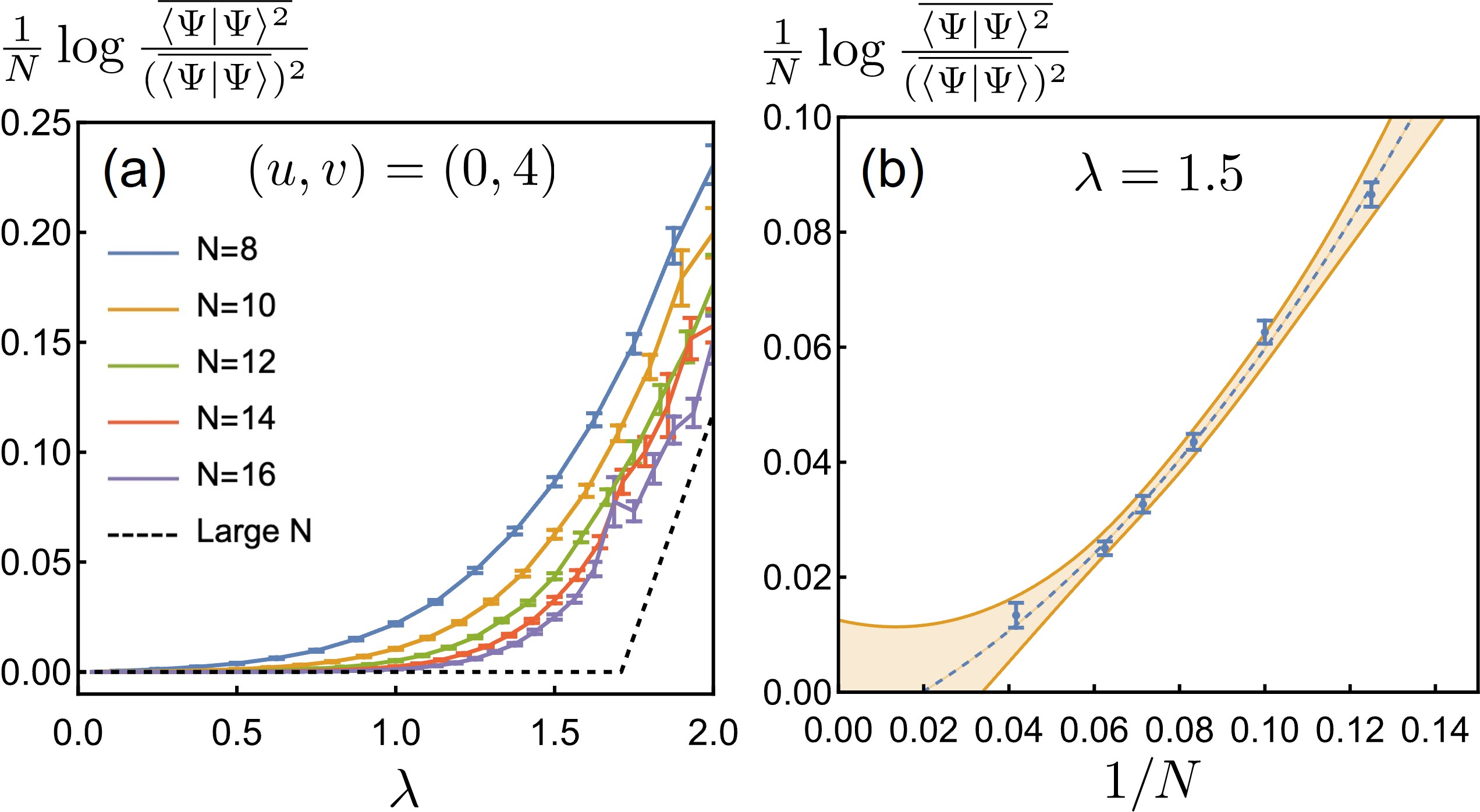}
    \caption{Evidence for a first-order transition in the squared norm fluctuations using finite-$N$ scaling.  (a) Fluctuations in the norm as characterized by $ \frac{1}{N} \log\frac{\overline{\braket{\Psi}{\Psi}^2}}{\left(\overline{\braket{\Psi}{\Psi}}\right)^2}$. $5\times 10^4$ random samples were used to compute the averages for $N=8,10,12,14,16$. The numeric results approach the analytic result (dashed line) at large $N$. (b) Finite-size scaling at $\lambda=1.5$ indicates that $\frac{1}{N}\log\frac{\overline{\braket{\Psi}{\Psi}^2}}{(\overline{\braket{\Psi}{\Psi}})^2}$ decreases towards zero (the large-$N$ prediction) as $N$ increases ($1/N\rightarrow 0$). The dashed line is the error weighted quadratic fit and the shaded area is the $90\%$ confidence band. To better access the large-$N$ behavior, we have added an additional data point using $10^4$ random samples of $N=24$.}
    \label{fig: norm_fluctuations}
\end{figure}

Finally, we compare our analysis with numerics at finite $N$. To verify the difference in the norm fluctuations between the two phases, we compute $\frac{1}{N} \log\frac{\overline{\braket{\Psi}{\Psi}^2}}{\left(\overline{\braket{\Psi}{\Psi}}\right)^2}$ as a function of $\lambda = M/N$ for different $N$. For $(u,v)=(0,4)$, we expect a first-order phase transition at $\lambda=\lambda_c\approx 1.71$ that should manifest as a non-differentiable change in the squared norm fluctuations. From Fig.~\ref{fig: norm_fluctuations}, one sees that numerics indeed show evidence of a first-order transition in the large-$N$ thermodynamic limit. 

\section{Second R\'{e}nyi entropy and phase transitions}
\label{sec:S2}
In this section, we examine the second R\'{e}nyi entropy of a generic subregion $A$ of the Gaussian RBM ensemble. Like the previous calculation of $\overline{\braket{\Psi}{\Psi}}$, we shall map the computation of the second R\'{e}nyi entropy to a statistical mechanics model. The relevant parameters of the statistical model are the subregion size $a\equiv \abs{A}/N$ as well as $(u,v,\lambda)$, which span a four-dimensional parameter space. For clarity, we will focus on the subspaces (i) $a=1/2$ for general $u,v$ and (ii) $(u,v)=(0,\infty)$ for general $a$, which will encompass the region of near-maximal entropy. In the second case, we shall analytically compute the second R\'{e}nyi entropy as a function of $a$ in the thermodynamic limit, and compare these to finite-$N$ numerics. In the following, we will refer to the curve of entropy versus $a$ as the ``Page curve''\cite{Page:1993} following the naming convention in the study of random states and black hole evaporation. Built upon the understanding of these special cases, we will later discuss qualitative features of Page curves for more generic parameters.

\subsection{Mapping to a statistical mechanics model}
Here, we introduce the analytic machinery for computing the second R\'{e}nyi entropy of the RBM ensemble in large-$N$ limit. For the parameter regime with vanishingly small norm fluctuations, the ensemble-averaged second R\'{e}nyi entropy may be computed exactly in the large-$N$ limit. We begin by expressing the second R\'{e}nyi entropy of a subregion $A$ using a doubled copy of the system $\ketbra{\Psi}{\Psi}$: 
\begin{equation}
    \label{eq: renyi}
    S_2(A) = -\log\left(\frac{\tr(X_{A}\ketbra{\Psi}{\Psi}\otimes \ketbra{\Psi}{\Psi})}{\tr(\ketbra{\Psi}{\Psi}\otimes \ketbra{\Psi}{\Psi})}\right),
\end{equation}
where $X_A$ is the swap operator acting to swap the subregion $A$ configurations of the two copies.  A proof of this
formula may be found in Ref.~\cite{Hastings:2010}. Note that the denominator of the expression in parentheses is equivalent to $Z_0$. Similarly, we shall define $Z_1$ to be the numerator,
\begin{equation}
\begin{split}
    Z_1 &\equiv \tr \left( X_A\ketbra{\Psi}{\Psi}\otimes \ketbra{\Psi}{\Psi}\right).
\end{split}
\end{equation}
The ensemble-averaged $S_2(A)$ may be expanded in powers of the fluctuations $\delta Z_1=Z_1-\overline{Z_1}$ and $\delta Z_0=Z_0-\overline{Z_0}$:
\begin{align}
\overline{S_2(A)}&=-\overline{\log \left(\frac{\overline{Z_1}+\delta Z_1}{\overline{Z_0}+\delta Z_0}\right)} \nonumber\\
&= -\log \frac{\overline{Z_1}}{\overline{Z_0}} + \sum_{n=1}^\infty \frac{(-1)^{n-1}}{n}\left(\frac{\overline{\delta Z_0^n}}{\overline{Z_0}^n} - \frac{\overline{\delta Z_1^n}}{\overline{Z_1}^n} \right) \nonumber \\
&\approx -\log \frac{\overline{Z_1}}{\overline{Z_0}}. \label{eq: s2_approx_ratio}
\end{align}
The approximation in Eq.~\eqref{eq: s2_approx_ratio} is only justified when the fluctuations $\delta Z_1$ and $\delta Z_0$ are small in the large-$N$ expansion, i.e., $\delta Z_0/Z_0\ll 1$, $\delta Z_1/Z_1 \ll 1$ as $N\rightarrow \infty$. Hence, the following analytic computation of the approximation $-\log\left(\overline{Z_1}/\overline{Z_0}\right)$ will only be applicable in a certain region of the parameter space. However, as we will demonstrate later, the approximation provides valuable insights into the behavior of entropy, and remains valid in the near-maximal entropy regime.

\begin{figure}
    \centering
    \includegraphics[width=.95\linewidth]{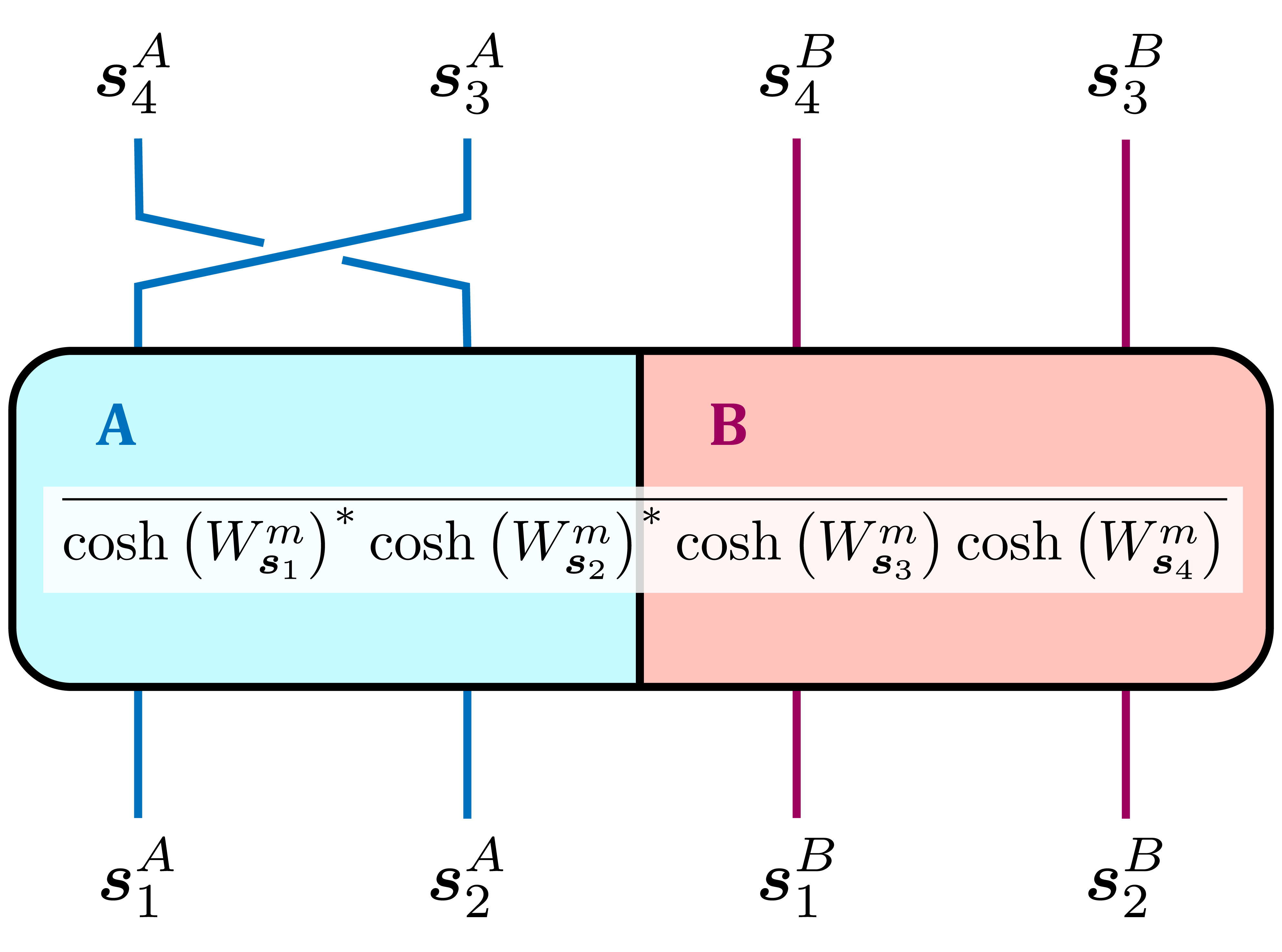}
    \caption{An illustration of ensemble-averaged tensor $X_{A}\overline{|\Psi\rangle\langle\Psi|\otimes |\Psi\rangle\langle\Psi|}$. The trace of this tensor gives $\overline{Z_1}$ and the contribution to the trace comes from the components which satisfy $\bs_{1}^{A}=\bs_{4}^{A}$, $\bs_{3}^{A}=\bs_{2}^{A}$, $\bs_{1}^{B}=\bs_{3}^{B}$ and $\bs_{2}^{B}=\bs_{4}^{B}$. Under these constraints, $\bs_1\cdot \bs_3=\bs_{2}\cdot \bs_{4}=N\phi_A+|B|$, $\bs_{1}\cdot \bs_{4}=\bs_{2}\cdot \bs_{3}=N\phi_B+|A|$ and $\bs_1\cdot \bs_2=\bs_3\cdot \bs_4=N(\phi_A+\phi_B)$.}
    \label{fig: Z1_tensor}
\end{figure}

We begin by computing the components of the ensemble-averaged tensor
\begin{equation}
\begin{split}
    &\overline{|\Psi\rangle \langle \Psi|\otimes|\Psi\rangle \langle\Psi|}=\sum_{\bs_1,\bs_2,\bs_3,\bs_4}\Gamma_{\bs_1\bs_2\bs_3\bs_4}|\bs_{3}\rangle\langle\bs_{1}|\otimes |\bs_{4}\rangle\langle\bs_{2}|,
\end{split}
\end{equation}
where
\begin{equation}
\resizebox{.95\hsize}{!}{$\Gamma_{\bs_1\bs_2\bs_3\bs_4}\equiv \displaystyle\prod_{m}\overline{\cosh(W_{\bs_1}^{m})^{*}\cosh(W_{\bs_2}^{m})^{*}\cosh(W_{\bs_3}^{m})\cosh(W_{\bs_4}^{m})}$}.
\end{equation}
Note that the trace of this tensor yields $\overline{Z_0}$ while the trace of the swap operator acting on the above tensor yields $\overline{Z_1}$, as shown in \figref{fig: Z1_tensor}. One can partition the spin configurations into the configurations on subsystem $A$ and its complement $B$
\begin{equation}
    \bs_k=(\bs_{k}^A,\bs_{k}^B),
\end{equation}
where $\bs_{k}^{A,B}\equiv\{s_k^{j_{A, B}}|j\in A, B\}$, and then express $\overline{Z_1}$ as
\begin{equation}
    \overline{Z_1}=\sum_{\bs_1 \bs_2}\Gamma_{\bs_1 \bs_2 \bs_3 \bs_4}\delta_{\bs_1^A\bs_4^A}\delta_{\bs_2^A\bs_3^A}\delta_{\bs_1^B\bs_3^B}\delta_{\bs_2^B\bs_4^B}.
\label{eq: Z1_tensor}
\end{equation}
The swap operator switches $\bs_3^A$ and $\bs_4^A$ while the trace identifies $\bs_1^A=\bs_4^A$, $\bs_2^A=\bs_3^A$, $\bs_1^B=\bs_3^B$ and $\bs_2^B=\bs_4^B$.

As with $\overline{Z_0}$, symmetry constrains the possible form of the expression for $\overline{Z_1}$ after performing the sum. Recall that the RBM ensemble is invariant under permutation or inversion of the random weights. This implies that $\Gamma_{\bs_1 \bs_2 \bs_3 \bs_4}$ is invariant under permutations or negations of the spin variables provided that the actions are performed in the four copies simultaneously. This restricts $\Gamma_{\bs_1 \bs_2 \bs_3 \bs_4}$ to be a function of $\bs_j\cdot \bs_k$ for $j$, $k$ $\in  1\dots 4$. After the identification implemented by the trace, $\Gamma_{\bs_1 \bs_2 \bs_3 \bs_4}$ can only depend on $\bs_1$ and $\bs_2$ through
two "order parameters"
\begin{equation}
    \phi_A\equiv \frac{1}{N}\sum_{j\in A}s_1^{j}s_2^{j},\qquad \phi_B\equiv \frac{1}{N}\sum_{j\in B}s_1^{j}s_2^{j}.
\end{equation}

One can further recast the expression for $\overline{Z_1}$ as a statistical mechanics model that depends on $\phi_A$ and $\phi_B$. In particular, $\overline{Z_1}$ becomes a partition function associated to coupled spin chains. The explicit formula for $\overline{Z_1}$ can be straightforwardly worked out using Gaussian integrals (see Appendix~\ref{app: Z1_computation}), and yields
\begin{equation}
    \overline{Z_1}=\sum_{\phi_A,\phi_B}\exp\left[-N \mathcal{F}(\phi_A,\phi_B)\right],
    \label{eq: spin_sum}
\end{equation}
where the free energy density $\SF$ is defined to be a difference of two contributions
\begin{equation}
    \mathcal{F}(\phi_A,\phi_B)=\mathcal{E}(\phi_A,\phi_B)-\mathcal{S}(\phi_A,\phi_B).
    \label{eq: Z1_free_energy}
\end{equation}
As before, $\CE$ represents the "energetic" contribution from the coupling of the spin chains,
\begin{equation}
\begin{split}
    \CE =&-2 \lambda(u^2-v^2)+\lambda\log 2\\
    &-\lambda\log\Big{[}1+\frac{e^{2\alpha}}{2}\cosh \Big(2(\beta+\gamma)\Big)\\
    &+\frac{e^{-2\alpha}}{2}\cosh\Big( 2(\beta-\gamma)\Big)\Big{]}
\end{split}
\label{eq: energy_AB}
\end{equation}
with
\begin{equation}
\begin{split}
   \alpha&\equiv(b+\phi_A)(u^2+v^2),\\
   \beta&\equiv(a+\phi_B)(u^2+v^2),\\
   \gamma&\equiv(\phi_A+\phi_B)(u^2-v^2).
\end{split}
\end{equation}
Meanwhile, $\CS(\phi_A,\phi_B)$ represents the "entropic" contribution from the degeneracy of spin configurations pairs ($\bs_1$,$\bs_2$) associated to a fixed ($\phi_A$, $\phi_B$),
\begin{equation}
    \mathcal{S}(\phi_A,\phi_B)=\log 2+ a H_b\left(\frac{a-\phi_A}{2a}\right)+b H_b\left(\frac{b-\phi_B}{2b} \right),
\label{eq: entropy_AB}
\end{equation}
where $a$ and $b$ represent the fraction of $N$ sites comprising subsystem $A$ and $B$, respectively ($a \equiv |A|/N$, $b \equiv |B|/N = 1-a$), and $H_b$ is the binary entropy function $H_b(p)\equiv -p\log p-(1-p)\log p$. The $\log 2$ comes from the spin-flip symmetry of $\phi_A$ and $\phi_B$.

Notice that $\overline{Z_1}$ reduces to $\overline{Z_0}$ if one sets $a=\phi_A=0$. For small fluctuations of $Z_0$ and $Z_1$, the second R\'{e}nyi entropy may be approximated by
\begin{equation}
    \frac{\overline{S_2}(A)}{N}\approx \mathcal{F}(\phi_A^{\star},\phi_B^{\star})+\frac{\log \overline{Z_0}}{N} ,
\label{eq: s2_approx}
\end{equation}
where $\mathcal{F}(\phi_A^{\star},\phi_B^{\star})$ is the global minimum of the free energy density and $(\phi_A^{\star},\phi_B^{\star})$ is its location. This approximation is accurate to lowest order in $1/N$ if the fluctuations in $Z_0$ and $Z_1$ are small.

\subsection{Half-system second R\'{e}nyi entropy}
\begin{figure}[t]
    \centering
    \includegraphics[width=\linewidth]{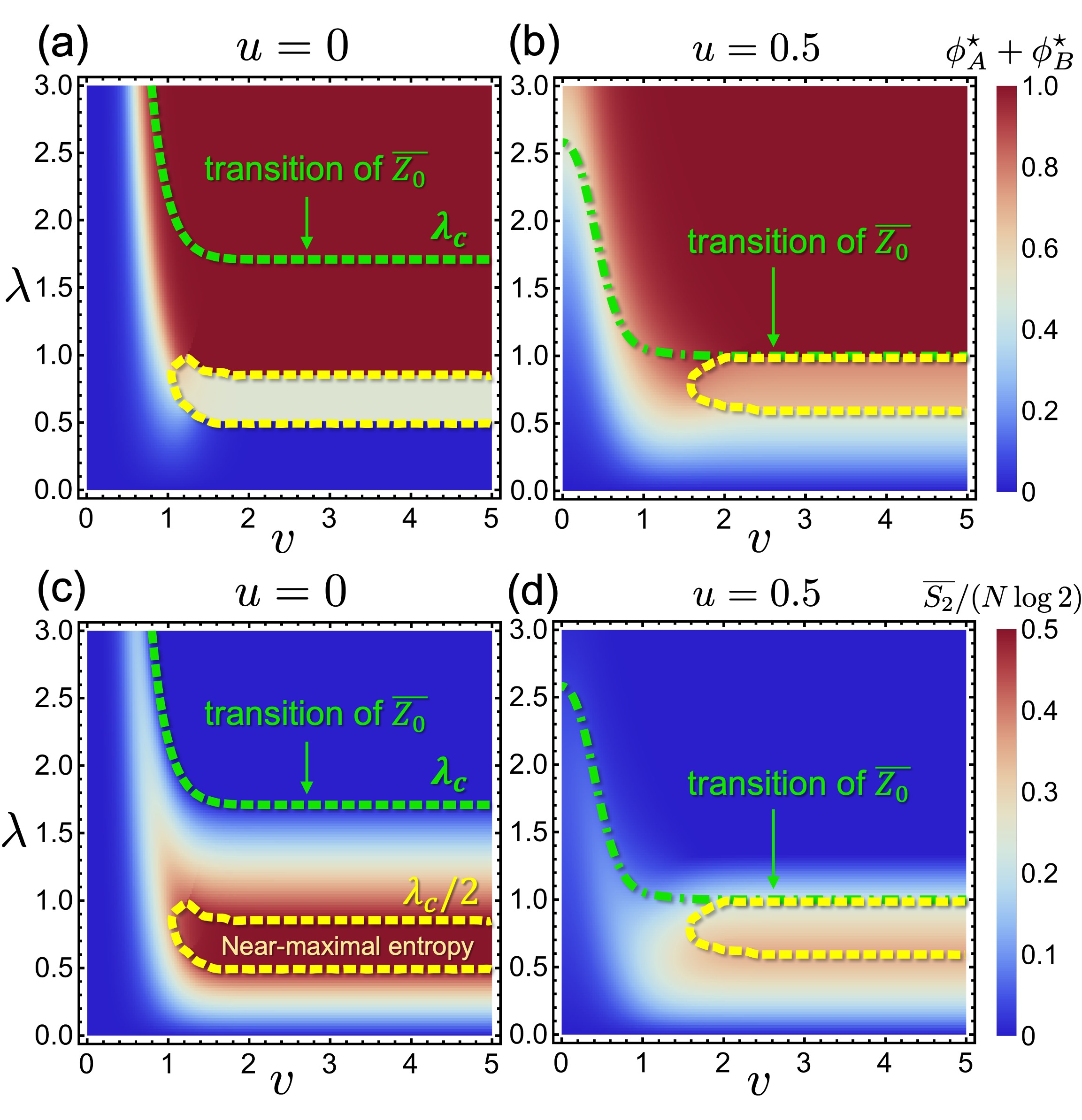}
    \caption{Phase diagrams for a half-system partition at fixed $u = 0$ (a, c) and $u = 0.5$ (b, d). Panels (a) and (b) capture information about the configuration $(\phi_A^{\star}, \phi_B^{\star})$ that minimizes the free energy density associated to $\overline{Z_1}$. Panels (c) and (d) plot the approximate averaged second R\'{e}nyi entropy $\overline{S_2(A)}$ using Eq.~\eqref{eq: s2_approx}. We point out that the phase diagram is unreliable above the transition line of $\overline{Z_0}$ which is when the analytic approximation in Eq.~\eqref{eq: s2_approx} breaks down due to large fluctuations in the squared norm. The dashed yellow line separates two phases of the free energy of $\overline{Z_1}$. The phase outside the dashed yellow line preserves the $\phi_A\leftrightarrow \phi_B$ symmetry, while the phase within the dashed yellow line breaks the symmetry. Interestingly, the symmetry-broken phase has near-maximal entropy for $(u,v)=(0,\infty)$ in the large-$N$ limit. For these parameters, the phase boundary is at $\lambda=1/2$ and $\lambda=\lambda_c/2$, where $\lambda_c$ is the critical $\lambda$ for the transition of $\overline{Z_0}$.}
    \label{fig: phase_diagram_S2}
\end{figure}

We begin by examining the half-system second R\'{e}nyi entropy ($a=b=1/2$), as it provides a useful measure for the degree of entanglement in a wavefunction. The hyperplanes $u=0, 0.5$ were selected for studying $\overline{S_2}$ as a function of $(v,\lambda)$. The phase diagrams in Fig.~\ref{fig: phase_diagram_S2} were obtained by numerically minimizing the free energy in Eq.~\eqref{eq: Z1_free_energy} and using the minimizing order parameters $(\phi_A^{\star},\phi_B^{\star})$ to compute the second R\'{e}nyi entropy~\footnote{To properly minimize the free energy and find the global minimum, one should properly take into account possible local minima near the $\phi_A=1/2$ or $\phi_B=1/2$ line and the corner $\phi_A=\phi_B=1/2$.}. Noticeably, since R\'{e}nyi entropy calculations of a subregion and its complement should give the same result, the free energy is symmetric under $a\leftrightarrow b$ and $\phi_A\leftrightarrow\phi_B$. Therefore, at $a=b=1/2$, the free energy is symmetric under $\phi_A\leftrightarrow\phi_B$. Accordingly, there are two phases which are classified by this symmetry. In the plots of $\phi_A^{\star}+\phi_B^{\star}$ [Fig.~\ref{fig: phase_diagram_S2}(a-b)], the blue regime is dominated by a free energy minimum $(\phi_A^{\star},\phi_B^{\star})$ close to $(0,0)$, while the dark red regime is dominated by a free energy minimum $(\phi_A^{\star},\phi_B^{\star})$ close to $\left(\frac{1}{2},\frac{1}{2}\right)$. Both regimes belong to the same phase since one can traverse the parameter space from one regime to the other without breaking the $\phi_A\leftrightarrow\phi_B$ symmetry. However, there does exist a phase that breaks the $\phi_A\leftrightarrow\phi_B$ symmetry, which we numerically found to be separated by a first order phase boundary (dashed yellow in Fig.~\ref{fig: phase_diagram_S2}). Nevertheless, there is in principle no reason to forbid the appearance of a second-order phase transition in other parameter regimes. 

Importantly, in the phase diagrams for both $u=0,0.5$, the boundary lines separating the symmetry-broken phase become independent of $v$ in the large-$v$ limit. In the specific case where $u=0$, the symmetry-broken phase can approach the maximal half-system entropy of $(N/2)\log 2$ at large $v$ to lowest order in $1/N$, as shown in Fig.~\ref{fig: phase_diagram_S2}(a,c). This is in contrast with the numerical investigation of a different RBM ensemble in Ref.~\cite{Deng:2017} where the maximal entropy is not reached. Note that the symmetry-broken phase resides fully in the regime of small norm fluctuations, which are required for our approximation of Eq.~\eqref{eq: s2_approx} to be accurate. We will confirm the accuracy of the analytic treatment in the near-maximal entropy regime using numerics, which are presented in the next subsection.  
\subsection{Page curves at the limit of $(u,v)=(0,\infty)$}
\label{subseq: page_curves}
We shall now focus on the limit of $u=0$ and $v=\infty$, where the maximal entropy can be achieved in the large-$N$ limit for a range of $\lambda$. To gain insight into the behavior of the second R\'{e}nyi entropy, we shall produce Page curves of $\overline{S_2(A)}$ by sweeping the value of $a$ between $0$ and $1$.

\begin{figure}[h]
    \centering
    \includegraphics[width=\linewidth]{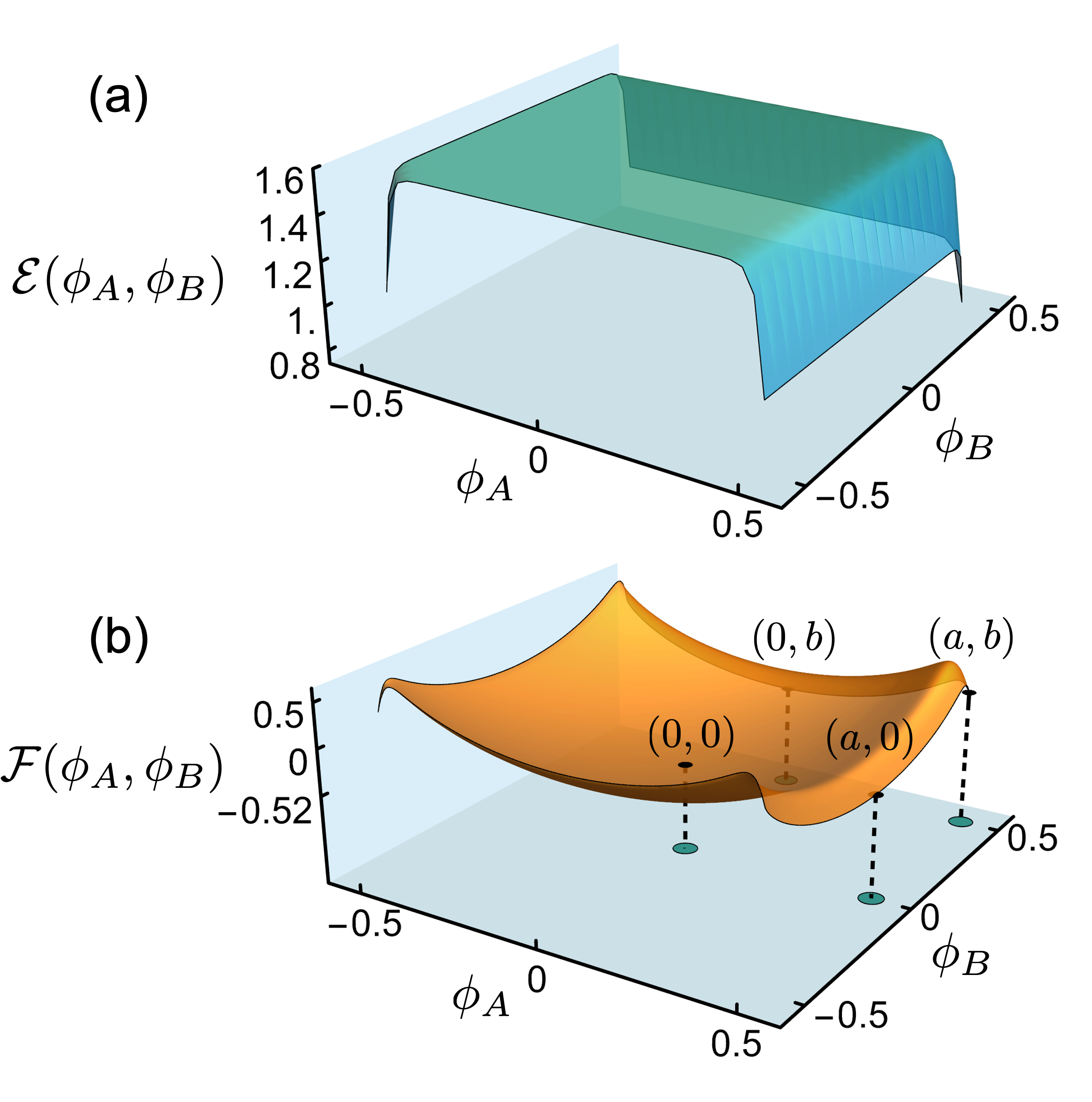}
    \caption{Surface plots of the (a) energy density and the (b) free energy density associated to $\overline{Z_1}$ for parameters $(a,u,v,\lambda)=(\frac{1}{2},0,4,\frac{3}{4})$. The energy density is sharply sloped near the boundaries of the $(\phi_A,\phi_B)$ domain, indicating that its behavior may be approximated by step functions. To high accuracy, the free energy density has four local minima at coordinates $(0,0)$, $(a,0)$, $(0,b)$, and $(a,b)$.}
    \label{fig: free_energy_minima}
\end{figure}

Let us study the free energy as a function of $(\phi_A,\phi_B)$ over the full domain $\mathcal{D}\equiv[-a,a]\times[-b,b]$. The dominating exponentials in Eq.~\eqref{eq: energy_AB} lead to step-like behavior near the edges of the domain, in analogy with the behavior of the energy density of $\overline{Z_0}$. In particular, we expect a near-constant plateau for $(\phi_A,\phi_B)$ on the interior of $\mathcal{D}$, which dips sharply on the boundary of $\mathcal{D}$ ($\phi_A=a$ or $\phi_B=b$) as well as at the corners $(\phi_A,\phi_B)=(a,b)$ or $(\phi_A,\phi_B)=-(a,b)$ [see Fig.~\ref{fig: free_energy_minima}(a) for an example at large but finite $v$]. Further taking into account the entropy density, we find the global minimum of the free energy is located at one of the four local minima $(\phi_A,\phi_B)=\{(0,0),(a,0),(0,b),(a,b)\}$. The corresponding free energy densities of these local minima are:
\begin{equation}
    \begin{cases}
    \mathcal{F}(0,0)&=-2\log 2,\\
    \mathcal{F}(a,0)&=-\lambda\log 2-\log 2-b\log 2,\\
    \mathcal{F}(0,b)&=-\lambda\log 2-\log 2-a\log 2, \\
    \mathcal{F}(a,b)&=-\lambda\log 3-\log 2.
    \end{cases}
    \label{eq:free_energy}
\end{equation}
Fig.~\ref{fig: free_energy_minima} shows that the situation at finite $v$ quickly approaches the $v\rightarrow \infty$ limit. Using parameters $(a,u,v,\lambda) = (1/2,0,4,3/4)$, we see that the energy density in Fig.~\ref{fig: free_energy_minima}(a) exhibits the predicted step-like function behavior near the edges of the domain, and the free energy in Fig.~\ref{fig: free_energy_minima}(b) possesses local minima at the expected locations to great accuracy.

\begin{figure}[b]
    \centering
    \includegraphics[width=\linewidth]{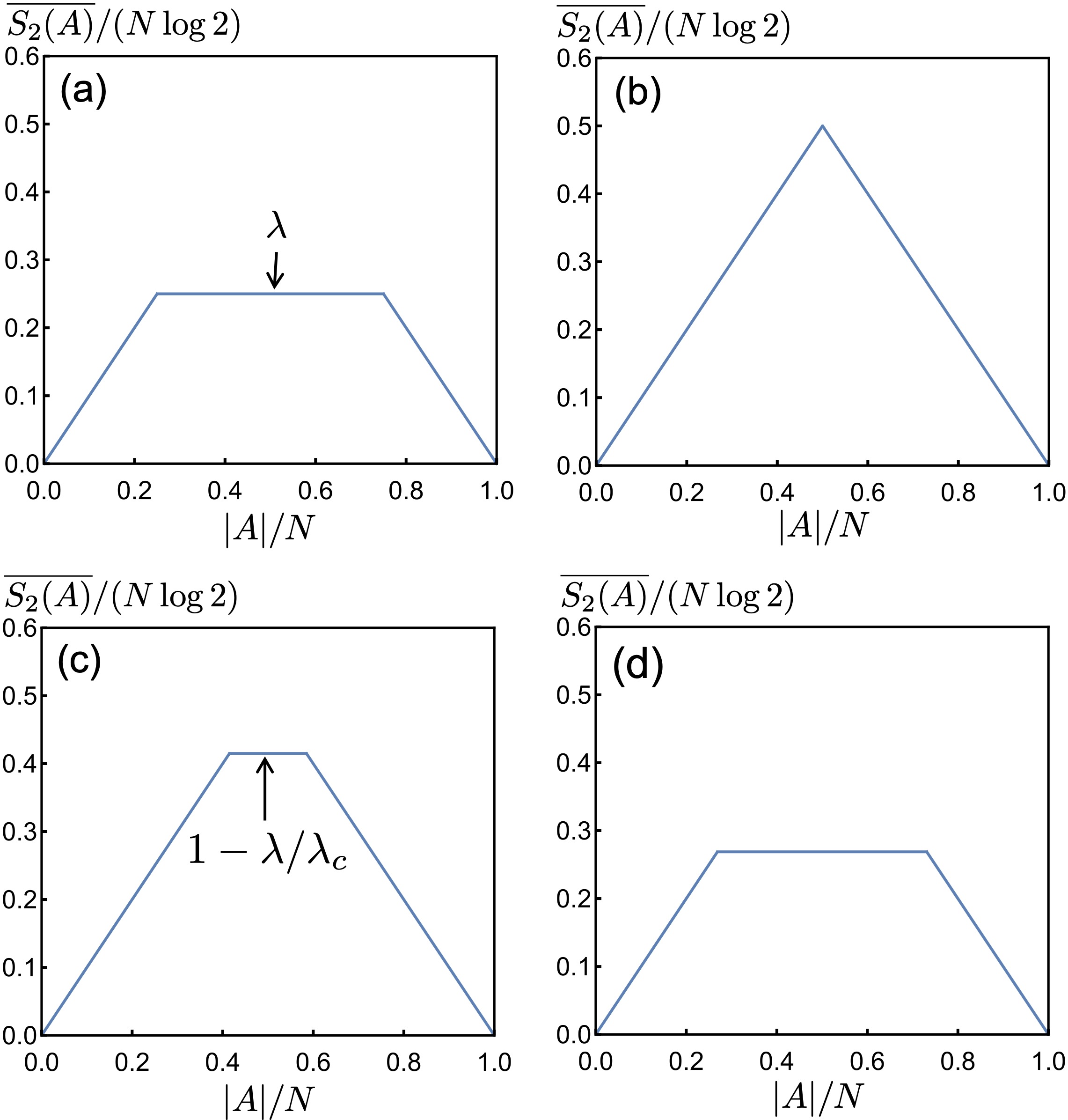}
    \caption{The Page curves for the averaged R\'{e}nyi entropy in the $(u,v)=(0,\infty)$ limit at $\lambda=0.25,0.75,1,1.25$ for panels (a-d). We can compare these results to Fig.~\ref{fig: phase_diagram_1}. For $\lambda\geq\lambda_c$, the norm fluctuations become large and $\overline{Z_0}$ will by a different free energy minimum at $|\phi|=1$. At this point, the estimate of $\overline{S_2(A)}$ from Eq.~\eqref{eq: s2_approx_ratio} becomes 0, reflecting a breakdown of the analytic estimate.}
    \label{fig: analytic_Page_curves}
\end{figure}

We may analytically construct the Page curve at different $\lambda$ by considering which of the free energy minima dominates for different $a$. Note that the approximation for $\overline{S_2(A)}$ in Eq. \eqref{eq: s2_approx} requires the norm fluctuations to be small, and as per the discussion in Sec.~\ref{sec:norm_fluc}, is strictly valid only for $\lambda$ below $\lambda_c$. First, for $a$ near $0$, one can compare free energies in Eq.~\eqref{eq:free_energy} and find that the global minimum of the free energy is at $(a,0)$. Hence, near the edge of the Page curve, the minimum free energy grows linearly as a function of $a$ and one can obtain the corresponding second R\'{e}nyi entropy as shown in the Page curves in Fig.~\ref{fig: analytic_Page_curves}. This suggests that in the chosen limiting case, the entropy saturates the maximal volume law for small subregions. Similarly, we anticipate analogous linearly decreasing behavior near $a=1$ because the Page curve is symmetric due to the spin permutation symmetry of the wavefunction ensemble.

When $(a,0)$ is no longer the global minimum of the free energy, the entropy with respect to subregion size transitions at some critical value $a=a_c$ from linear growth to a plateau, as shown in Fig.~\ref{fig: analytic_Page_curves}(a,c,d). In this regime, as the subregion size increases, the entropy does not obey a volume law. Due to the symmetry of the Page curve, the plateau will terminate at $a=1-a_c$. The dominant minimum in the plateau regime of the Page curve as well as the width of the plateau depends on the value of $\lambda$. Notably, if $a_c=1/2$, there is no plateau regime, but rather a direct transition between the two linear regions. One can solve for the global minimum in each of the regimes, and we summarize the results for different $\lambda$ below:
\begin{itemize}
\item For $\lambda<1/2$, $a_c=\lambda$ and the plateaued regime is dominated by the free energy minimum at $(\phi_A,\phi_B)=(0,0)$ yielding $\overline{S_2}/N\approx \lambda\log 2$. The plateau behavior can be understood as saturating the entropy bound set by the number of hidden neurons: $M\log 2$.
\item For $1/2<\lambda<\lambda_c/2$, there is no plateaued regime. The linear growth of the entropy reaches its maximal value at $a=1/2$, which also saturates the Page value of $\overline{S_2}/N =  \frac{1}{2} \log 2$. At $a=1/2$, there are degenerate free energy minima at $(\phi_A,\phi_B)=(1/2,0)$ and $(0,1/2)$. The large-$N$ solution spontaneously breaks the $\phi_A\leftrightarrow \phi_B$ symmetry, which defines the phase within the dashed yellow line in Fig.~\ref{fig: phase_diagram_S2}(a,c).
\item For $\lambda_c/2<\lambda<\lambda_c$, $a_c=1-\lambda/\lambda_c$ and the plateaued regime is dominated by the free energy minimum at $(\phi_A,\phi_B)=(a,b)$. 
\end{itemize}

\begin{figure}[htbp]
    \centering
    \includegraphics[width=\linewidth]{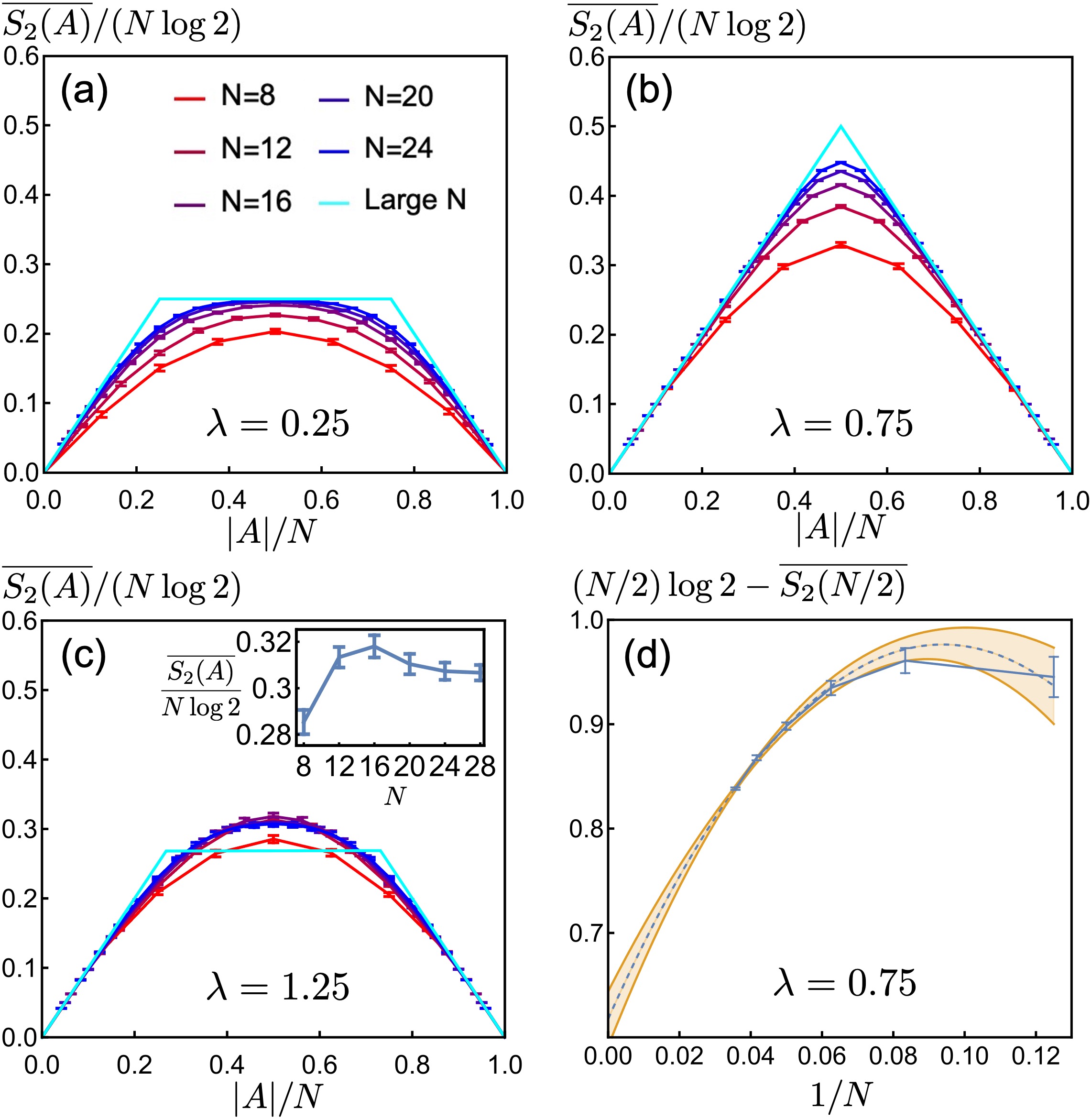}
    \caption{(a-c) Numerically-generated Page curves of the ensemble-averaged second R\'{e}nyi entropy for $(u,v)=(0,4)$ and $N=8,12,16,20,24$ (gradient coloring from red to blue), with $\lambda=0.25, 0.75, 1.25$ for panels (a), (b) an (c), respectively. Each ensemble-averaged point on the page curves was generated using 100 random samples. The analytic large-$N$ Page curves for $(u,v)=(0,\infty)$ are shown in cyan. The inset in panel (c) shows the details of the second R\`{e}nyi entropy at half-system size as a function of $N$ (including an additional point of $N=28$) at $\lambda=1.25$. While the numerically obtained Page curves converge to the analytic results in (a-b), there is no obvious convergence to the analytic result for $\lambda=1.25$ in (c). (d) The deficit between the the half-system second R\`{e}nyi entropy and the maximal value of $(N/2)\log 2$ at $\lambda=0.75$. The finite-size scaling trend of the deficit, which includes an additional point at $N=28$, suggests that at large $N$, the deficit approaches a value close to that of a generic random state ($\log 2\approx 0.69$). The dashed line is the error weighted quadratic fit and the shaded area is the $90\%$ confidence band.}
    \label{fig: numeric_Page_curves_s2}
\end{figure}

These analytic results are also summarized in Fig.~\ref{fig: analytic_Page_curves}. Interestingly, the analytics predicts that for large $\lambda$ (i.e. $\lambda>\lambda_c/2$), the height of the plateau decreases with increasing $\lambda$. We can  further compare the analytically determined Page curves with the finite-$N$ numerics, as shown in Fig.~\ref{fig: numeric_Page_curves_s2}. The numerical results are obtained using $(u,v)=(0,4)$, where the analytic calculations for $(u,v)=(0,\infty)$ can work sufficiently well (see Fig.~\ref{fig: free_energy_minima}).  
One can see that the numerically obtained Page curves converge to the analytic results for small $\lambda$ [see Fig.~\ref{fig: numeric_Page_curves_s2}(a-b)]. There is no obvious convergence to the analytic result at a large $\lambda$ of $1.25$ [see Fig.~\ref{fig: numeric_Page_curves_s2}(c)], which is not surprising since the analytic approximation in Eq.~\eqref{eq: s2_approx} breaks down when the fluctuation $\delta Z_0$ or $\delta Z_1$ becomes large as can be expected at the large $\lambda$ regime. The determination of the critical $\lambda$ that the breakdown occurs will be left for future studies. Nevertheless, the analytic computation does correctly capture the feature that the maximum half-system entropy is suppressed at large $\lambda$. We now offer an argument for the entropy suppression at large $\lambda$. Ref.~\cite{multifractality2020} shows that the entanglement entropy is bounded from above by the fractal dimension of the many-body wavefunction, which measures the degree of uniformity of the wavefunction amplitude in certain computational basis. As $\lambda$ increases, the wavefunction components become more non-uniform in the Ising basis, which limits the entanglement at large $\lambda$. More details about the fractal dimension and its calculation in our ensemble will be presented in Sec.~\ref{sec:entanglement_bound}. 

Remarkably, for $\lambda=0.75$---which is within the regime of near-maximal  second R\'{e}nyi entropy---we observe that the deficit in the maximum entropy $N/2\log 2-\overline{S_2}(N/2)$ appears to converge to some small $N$-independent value. The finite-size scaling plot in Fig.~\ref{fig: numeric_Page_curves_s2}(d) indicates that in the thermodynamic limit, the average second R\'{e}nyi entropy is nearly maximal, and the plotted deficit entropy is close to that of Haar-random states ($\log 2\approx 0.69$)~\cite{Lubkin:1993}. That being said, given that the RBM ensemble has only a polynomial number of free parameters in the system size, we expect that it will not be able to fully capture all features of the ensemble of Haar-random quantum states. We will show explicitly in Sec.~\ref{sec: other_measure} that different from Haar-random states, these states exhibit non-ergodic behavior in the Ising basis and do not form a quantum-state design.%\XQ{Need Revision: This motivates us to investigate more fine-grained diagnostics that can qualitatively distinguish between random RBM states in the near-maximal entropy regime and Haar-random states. We will postpone a more detailed discussion of this to Sec.~\ref{sec: other_measure}. }

\subsection{Possible types of Page curves}
\label{subsec: Page_types}
\begin{figure}[htbp]
    \centering
    \includegraphics[width=\linewidth]{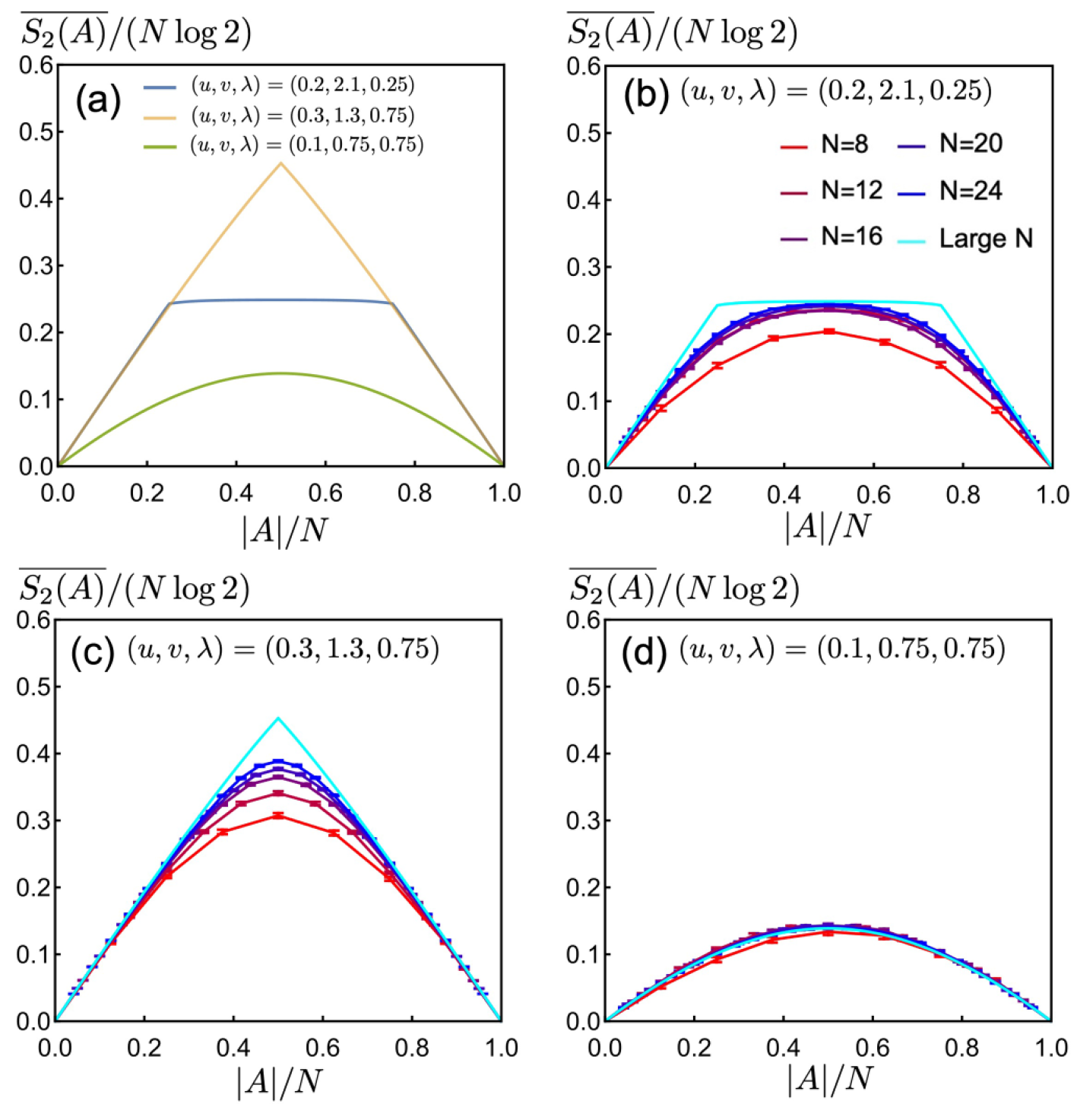}
    \caption{(a) Three qualitatively different types of Page curves from the large-$N$ calculation of the second R\'{e}nyi entropy at generic values of $(u,v,\lambda)$. (b-d). Numerically-generated Page curves averaged over $100$ RBM states in our ensemble at finite $N$ for the three parameter choices used in panel (a). The numeric curves consist of $N=8,12,16,20,24$ with a gradient color from red to blue.}
    \label{fig: general_page}
\end{figure}

Now considering the whole $(u,v,\lambda)$ parameter space, we may classify the different types of Page curves that can be obtained based on qualitative features that persist in the large-$N$ limit. Obtaining the full phase diagram for the Page curves requires solving the statistical mechanics problem in Eq.~\eqref{eq: Z1_free_energy} in the four-dimensional parameters space $(a,u,v,\lambda)$, and will not be the task of the present paper. Instead, we obtain Page curves from several representative points in parameter space, as shown in Fig.~\ref{fig: general_page}. Interestingly, we observe Page curves of three qualitatively different types (see Fig.~\ref{fig: general_page}), which can be classified based on the following features: (i) a shallow middle region and two non-differentiable corners, (ii) a single non-differentiable kink at $a=1/2$, and (iii) no corners or kinks. This indicates that the presence of non-differentiable kinks/corners is not restricted to the special case considered in the previous subsection, but is a more general feature which indicates a transition in the free energy minimum. Finite-$N$ numerics show strong support for the existence of Page curves of all three types for generic $(u,v,\lambda)$. Indeed, as shown in Fig.~\ref{fig: general_page}(b-d), the numeric Page curves converge to the analytic predictions as $N$ increases for these representative parameter points.  

\section{Other entanglement features}
\label{sec: other_measure}
In this section, we study other features of the RBM ensemble such as the von Neumann entropy, the entanglement spectrum and its level statistics, the fractal dimensions, and the relation of the ensemble to quantum state designs. These features will reveal further characteristics of the different regimes observed in the preceeding sections. In particular, these properties will clarify the similarities and differences between random RBM states and Haar-random quantum states in the regime of near-maximal entanglement entropy. We also explain some of our results using a bound on entropy from the fractal dimension. %\textcolor{blue}{These results can also serve as a guide in the search for entanglement transitions in related models} \cite{2021PhRvB.104j4205M}.

\subsection{Von Neumann entropy}
\label{subsec: SvN}
\begin{figure}[t]
    \centering
    \includegraphics[width=\linewidth]{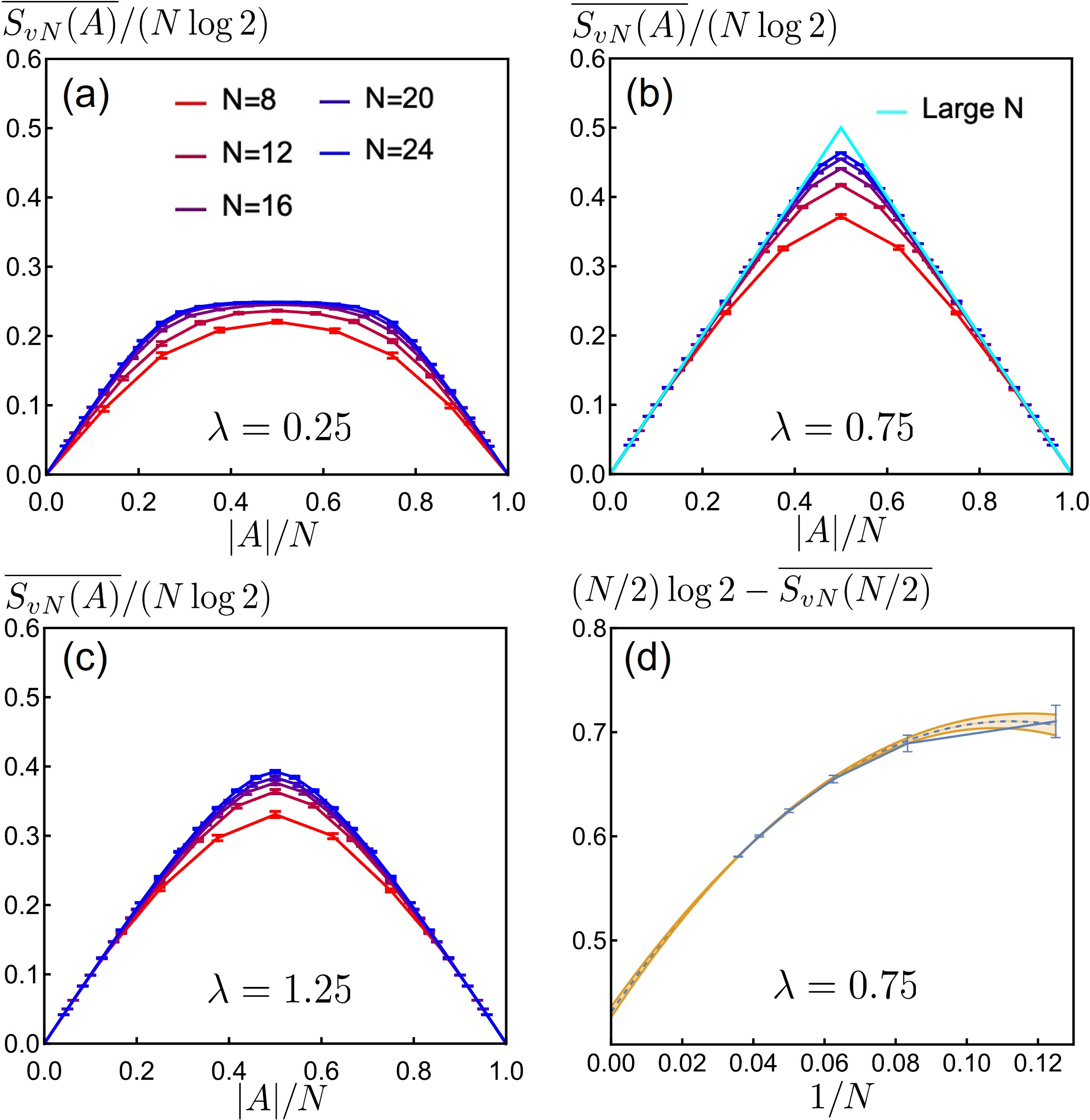}
    \caption{(a-c) Numerically-generated Page curves of the ensemble-averaged von Neumann entropy for $(u,v)=(0,4)$ and $N=8,12,16,20,24$ (gradient coloring from red to blue), with $\lambda=0.25, 0.75, 1.25$ for panels (a), (b) an (c), respectively. Each ensemble-averaged point on the page curves was generated using 100 random samples. Note that in (a), the entropy near the middle of the curve approaches a saturation value of $M \log 2$, just like the second R\'{e}nyi entropy in Fig.~\ref{fig: numeric_Page_curves_s2}(a) In (b), we compare the finite-$N$ curves in the near-maximal entanglement regime with the maximal entanglement Page curve (cyan). (d) Finite-size scaling of the deficit between the half-system von Neumann entropy and the maximal possible value of $(N/2)\log 2$ suggests that it approaches a constant that is close to the deficit for a random quantum state. Note that we have included an additional data point for $N=28$ averaged over $100$ random samples. The dashed line is the error weighted quadratic fit with the shaded area being the $90\%$ confidence band.}
    \label{fig: Page_curves_vN}
\end{figure}

The numeric study of the ensemble-averaged second R\'{e}nyi entropy can be straightforwardly extended to the von Neuman entropy. For comparison with the second R\'{e}nyi entropy, we use the same parameters as before, fixing $(u,v)=(0,4)$ and obtaining Page curves at different $\lambda$. The finite-$N$ Page curves of the ensemble-averaged von Neumann entropy are shown in Fig.~\ref{fig: Page_curves_vN}. Generally, one expects qualitatively similar behavior since the von Neumann entropy is an upper bound for the second R\'{e}nyi entropy. Indeed, one can see from Fig.~\ref{fig: Page_curves_vN} that the von Neumann entropy follows a trend similar to that of the second R\'{e}nyi entropy: near the edges of the Page curve, one observes near-maximal volume-law behavior, while the behavior near $a = \frac{1}{2}$ is strongly $\lambda$-dependent. The linear behavior at the edges is anticipated since the von Neumann entropy is bounded from below by the second Rényi entropy which has been shown to be near maximal.. Likewise, at small $\lambda$, the plateau-like behavior near the center of the curve has a value near the $M \log 2$ upper bound similar to the behavior of the second R\'{e}nyi entropy. For $\lambda = 0.75$ the von Neumann entropy approaches the maximal-entropy curve expected for a random quantum state to lowest order at large $N$. This is expected in the regime where one observes the maximal second R\'{e}nyi Page curve. Furthermore, like the second R\'{e}nyi entropy, a finite-size scaling analysis shows that the deficit between the half-system von Neumann entropy and the maximal possible value of $(N/2)\log 2$ approximately approaches a finite value, which is close to the theoretical deficit of $1/2$ for Haar-random states~\cite{Page:1993}. This represents yet another characteristic shared by random quantum states and RBM wavefunctions at these parameters. At larger $\lambda$, similar as the second R\'{e}nyi entropy, the von Neumann entropy is also suppressed and becomes sub-maximal.
\subsection{Entanglement spectrum and level statistics}
\label{subsec:level_stat}
The entanglement spectrum refers to the eigenvalue spectrum of $-\log \rho_A$ where $\rho_A$ is the reduced density matrix for subregion $A$. The entanglement spectrum probes more fine-grained entanglement properties of a quantum state than the entanglement entropy, and has been a powerful tool in characterizing quantum phases of matter~\cite{Li:2008}. In this subsection, we will study the entanglement spectrum, and its level statistics for the RBM ensemble. We will demonstrate that for the near-maximal entanglement regime, the density of states of the entanglement spectrum is close to the Marchenko-Pastur distribution for random states~\cite{Marchenko:1967,Marko:2006}. In contrast, for other regions of the phase diagram, we observe a deviation from the Marchenko-Pastur distribution, which is consistent with the results of a previous study of a slightly different ensemble~\cite{Deng:2017}. We will also show that the level statistics in the small $\lambda$ regime can be described by the random matrix results within a symmetry sector of our ensemble while deviations occur for large $\lambda$. 

\begin{figure}
    \centering
    \includegraphics[width=\linewidth]{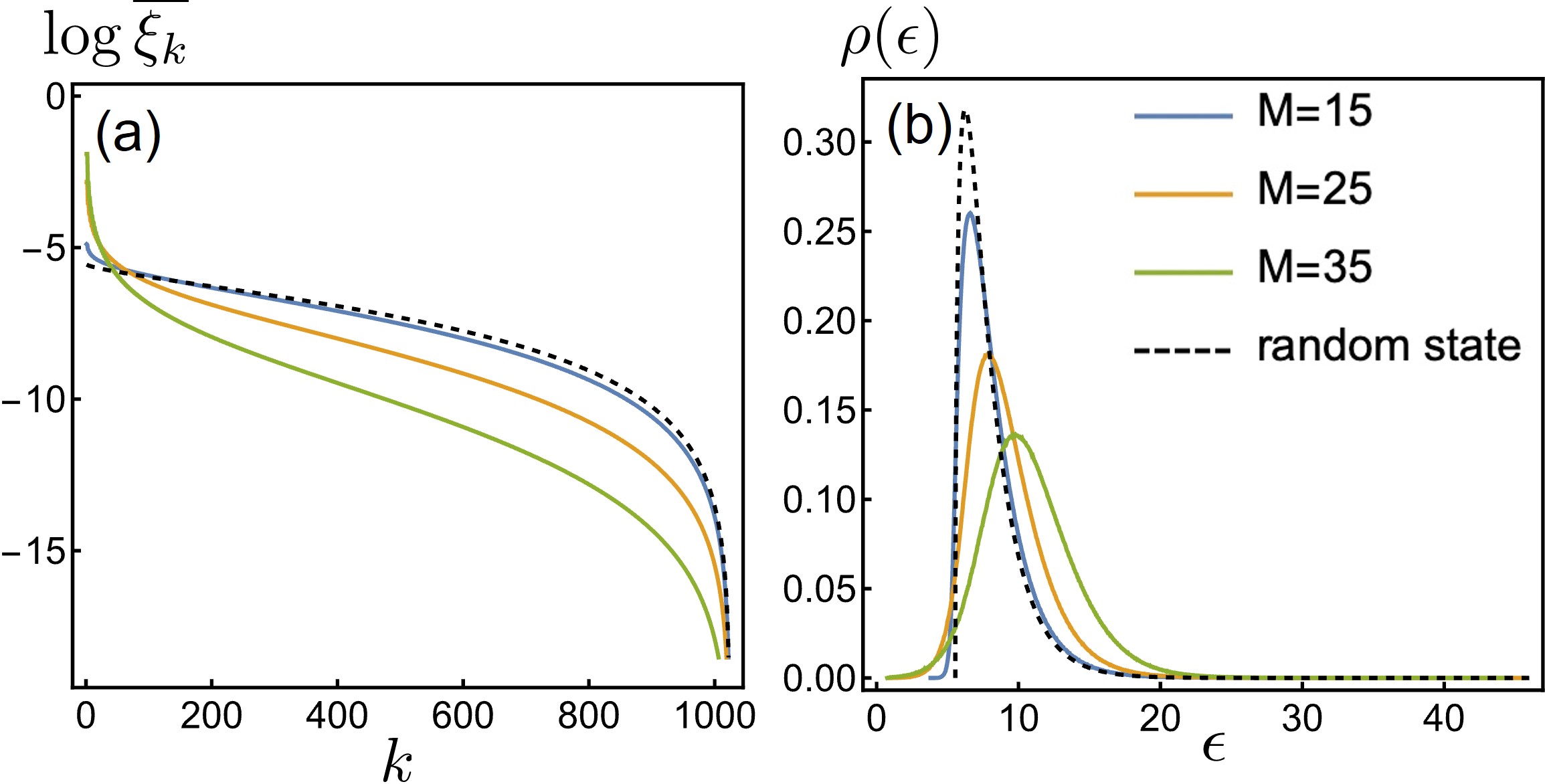}
    \caption{(a) The average entanglement spectrum for a half-system partition for $N=20$ at $(u,v)=(0,4)$ and for different numbers of hidden neurons: $M=15,25,35$ ($\lambda=0.75, 0.125, 0.175$). $\overline{\xi_{k}}$ is the ordered list of all eigenvalues of the reduced density matrix (arranged in descending order) upon averaging over $10^4$ randomly generated states in our ensemble. (b) The density of states of the entanglement Hamiltonian calculated from the histogram of sampled $\epsilon_k\equiv-\log \xi_k$ from all $10^4$ randomly generated states in our ensemble (area under the curve normalized to $1$). The dashed line is obtained from the Marchenko-Pastur distribution, which is associated with the Wishart random matrix ensemble for a random state.} %\TN{Should mention what scale parameter of the MP distribution was used for comparison.}}
    \label{fig:entanglement_spec}
\end{figure}

In the following,  to compare with previous discussions of $(u,v)=(0,\infty)$, we choose $N = 20$ and study the entanglement spectrum for a half-system bipartition ($|A|$ = 10) with parameters $(u,v)=(0,4)$ while varying $\lambda$. The average entanglement spectrum is obtained by taking the average of the list of eigenvalues $\xi_k$ of the reduced density matrix arranged in descending order. The density of states is obtained from appropriately normalized histograms of all sampled eigenvalues. The results are plotted in Fig.~\ref{fig:entanglement_spec}. Note that the rank of the reduced density matrix is bounded from above by $2^M$ [see discussion below Eq.~\eqref{eq: wf_product_states}]. Therefore for $\lambda<0.5$ (i.e. $M < 10 = |A|$), the reduced density matrix has at most $2^M$ nonzero eigenvalues. We avoid this reduction of nonzero eigenvalues by choosing $\lambda>0.5$; namely, we select $\lambda=0.75,1.25,1.75$ (corresponding to $M=15,25,35$).
For the study of the average entanglement spectrum and the density of states of the entanglement Hamiltonian [Fig.~\ref{fig:entanglement_spec}(a)], we do not show cases for $\lambda<0.5$, which certainly deviate from results of the Marchenko-Pastur distribution for a generic random state because of the presence of zero eigenvalues of the reduced density matrix. On the other hand, when studying the level statistics [Fig.~\ref{fig:entanglement_spec}(b)], we include a case where $\lambda<0.5$, but include only the non-zero eigenvalues of the reduced density matrix in counting the correlation between levels.

Interestingly, for $\lambda=0.75$ which is in the near-maximal entropy regime, the average entanglement spectrum and density of states are very close to those of a random state derived from Marchenko-Pastur distribution~\cite{Marchenko:1967,Marko:2006}, which is a novel result compared to the observations of Ref.~\cite{Deng:2017}. Since the regime has near-maximal entropy, we expect almost all of the degrees of freedom in subregion $A$ to be entangled and the entanglement spectrum to have $2^{|A|}$ nearly degenerate levels.  
However, higher order $1/N$ effects split the zeroth-order degenerate levels, producing a distribution close to the Marchenko-Pastur distribution for a random state.
%\XQ{Note that the approximate agreement of the entanglement spectrum for the near-maximal entropy RBM wavefunction and that of a random quantum spin state is a much stronger result than the saturation of the entropy bound. It also implies that all half-system R\'{e}nyi entropies approximately agree with those of a random state.}
%\XQ{Note that the approximate agreement of the entanglement spectrum for the near-maximal entropy RBM wavefunction and that of a random quantum spin state is a much stronger result than the saturation of the entropy bound. It also implies that all half-system R\'{e}nyi entropies approximately agree with those of a random state.}
While the entanglement spectrum for the near-maximal entropy RBM wavefunction mostly agrees with that of a random quantum spin state, we cannot yet conclude that all half-system R\'{e}nyi entropies also approximately agree with those of a random state. As we can see from Fig.~\ref{fig:entanglement_spec}(a), for the large eigenvalues of the reduced density matrix, there are some deviations and these can cause deviations in the higher R\'{e}nyi entropy. The behavior of these higher R\'{e}nyi entropies can be understood from a simple bound, which will be discussed in Sec.~\ref{sec:entanglement_bound}.

Notably absent from the above discussions of average entanglement spectrum and density of states is information about the statistical correlations \textit{between} levels in the spectrum, which is captured by the level statistics. Much like the level statistics of a \textit{Hamiltonian} spectrum, the level statistics of the \textit{entanglement} spectrum may be used to diagnose thermalization in quantum systems as demonstrated in Ref.~\cite{Geraedts:2016}. It was found that for a thermalized eigenstate, the level statistics of the entanglement spectrum are captured by the random matrix ensemble in the corresponding symmetry class of the associated Hamiltonian, namely, Gaussian orthogonal ensemble (GOE), Gaussian unitary ensemble (GUE) or Gaussian symplectic ensemble (GSE). On the other hand, for a many-body localized state, the level statistics of the entanglement spectrum follow a semi-Poisson distribution~\cite{Geraedts:2016} (which is different from the predicted distribution for the energy level statistics for an integrable system). 
Here, we will investigate the level statistics of the entanglement spectrum in different parameter regimes to characterize the types of wavefunctions encoded by our Gaussian RBM ensemble.

A previous study~\cite{Deng:2017} has reported Poisson-like level statistics for zero-bias RBM states with random weights. Our results confirm that the level statistics is Poissonian. However, we note that there is a simple symmetry reason for the Poissonian statistics. %\TN{However, we note that the lack of biases gives rise to a parity symmetry of the wavefunction components, which must be taken into account before studying the level statistics. 
Defining a global spin flip operator on the physical spins $\Sigma=\prod_{j}\sigma_x^j$ ($\sigma_x^j$ are Pauli matrices), we notice that $\Sigma\ket{\psi} = \ket{\psi}$. We can also write the symmetry operator as $\Sigma=\Sigma_A\Sigma_B$, where $\Sigma_A=\prod_{j\in A}\sigma_x^j$ and $\Sigma_B=\prod_{j\in B}\sigma_x^j$ ($B$ is the complement of subregion $A$). This produces a symmetry in the reduced density matrix $\rho_A$, which allows it to be block diagonalized: $\rho_A=\tr_B(|\psi\rangle\langle \psi|)=\tr_B(\Sigma_A\Sigma_B|\psi\rangle\langle \psi|\Sigma_B\Sigma_A)=\Sigma_A \rho_A \Sigma_A$. Thus, the entanglement levels have $\Sigma_A$ eigenvalues of $\pm 1$ and the absence of the mixing of the levels in different symmetry sectors can lead to the previously reported Poisson-type behavior of zero-bias random RBM states. In the following, we study the level statistics within one symmetry sector of $\Sigma_A=+1$.

\begin{figure}[htbp]
    \centering
    \includegraphics[width=\linewidth]{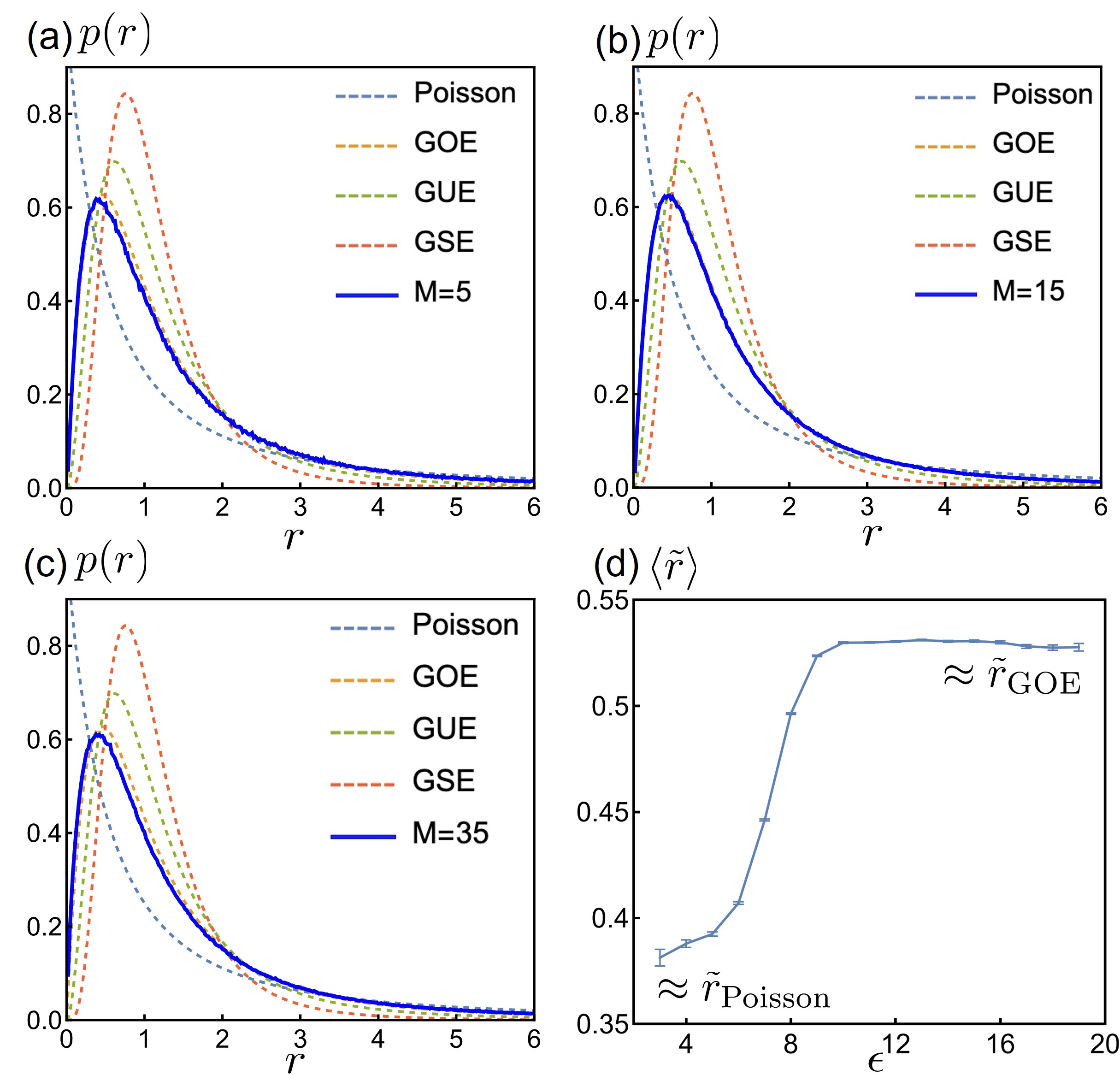}
    \caption{Distribution of level spacing ratios $r$ of the entanglement spectrum for $N = 20$ and $M=5,15,35$ in panels (a-c), respectively. These are plotted against the Poisson and Gaussian ensemble distributions for comparison. The entanglement spectra are obtained for half subsystem and parameters $(u,v)=(0,4)$. The empirical distribution is obtained from a normalized histogram over $10^4$ random samples, each with $2^{10}$ eigenvalues of the corresponding entanglement Hamiltonian for $M=15,35$. We take half of the eigenvalues corresponding to symmetry sector $\Sigma_A=+1$. The bin width used to generate these plots is $0.02$. For $M=5$, we obtain the empirical distribution from a normalized histogram of the $16$ nonzero eigenvalues of the reduced density matrix over $10^5$ random samples in the symmetry sector $\Sigma_A=+1$. Panel (d) shows the averaged reduced level spacing ratio calculated for eigenvalues of the entanglement Hamiltonian within a range $(\epsilon-\Delta\epsilon,\epsilon+\Delta\epsilon)$ for $M=35$ and $(u,v)=(0,4)$. We choose $\Delta\epsilon=0.5$. The results indicate that for large $\lambda$ small eigenvalues of the entanglement Hamiltonian $-\log \rho_A$ (large eigenvalues of $\rho_A$) have level statistics close to Poisson.}
    \label{fig:levelstats}
\end{figure}

We adopt the formalism of Ref.~\cite{Oganesyan:2007} for studying the level statistics. In particular, the $n^\text{th}$ level spacing ratio $r_n$ is defined to be
\begin{equation}
    r_n \equiv \delta_n/\delta_{n-1},
\end{equation}
where $\delta_n\equiv\epsilon_{n+1}-\epsilon_{n}$ is the difference between the $n^\text{th}$ level and $(n+1)^\text{th}$ level of the sorted entanglement spectrum $\{\epsilon_n;\epsilon_n\le \epsilon_{n+1}\}$ in the symmetry sector. The distribution of the level spacing ratios will inform us about the statistical correlations in the entanglement spectrum. %, and provide more fine-grained information for comparison of our RBM ensemble with the Haar-random ensemble. 
For direct comparison with our prior results, we select the same parameters $(a,u,v)=(\frac{1}{2},0,4)$ and collect entanglement spectrum data for randomly drawn Gaussian RBM states with a fixed $N=20$ and varying $M$. From the whole set of empirically-generated $r_n$, we can obtain the (approximate) distribution $p(r)$, which is the probability density that a level spacing ratio $r$ occurs. %The numeric results, which are summarized in Fig.~\ref{fig:levelstats}, indeed show level repulsion for \textit{any} of the explored values of $\lambda$ in the three regimes (small $\lambda$, intermediate $\lambda$ with near-maximal entropy behavior, and large $\lambda$). %\XQ{Note that for $\lambda=0.25$ ($M=5$), we consider only the $2^5/2 = 16$ nonzero eigenvalues (the factor of $1/2$ comes from choosing one symmetry sector) of the reduced density matrix when producing the distribution $p(r)$. 
The numeric results are summarized in Fig.~\ref{fig:levelstats}. In \textit{any} of the explored values of $\lambda$ in the three regimes (small $\lambda$, intermediate $\lambda$ with near-maximal entropy behavior, and large $\lambda$), the wavefunction is real, and the probability distribution $p(r)$ follows the GOE result for most values of $r$. However, at large $\lambda$, the deviation from the GOE result occurs close to $r=0$: the level spacing ratio has an increasingly nonvanishing probability of being zero, as can be seen in Fig.~\ref{fig:levelstats}(c). This becomes more clear when considering a spectrum-resolved level spacing ratio. To illustrate this, we consider the average of the reduced level spacing ratio $\tilde{r}_n \equiv \text{min}(r_n,1/r_n)$ within a window of $(\epsilon-\Delta\epsilon,\epsilon+\Delta\epsilon)$ in the entanglement spectrum. It turns out the average reduced level spacing ratio $\langle \tilde{r}\rangle$ is strongly dependent on the region of the spectrum, as shown in Fig.~\ref{fig:levelstats}(d). The results suggest that for small eigenvalues of the entanglement Hamiltonian (i.e. large eigenvalues of the reduced density matrix), the level statistics are close to Poisson ($\tilde{r}\approx 0.386$~\cite{Atas:2013}). For large eigenvalues of the entanglement Hamiltonian (i.e. small eigenvalues of the reduced density matrix), the level statistics gradually switch to the GOE result. We will later explain the behavior at large $\lambda$ using an analysis of fractal dimensions in Sec.~\ref{sec:entanglement_bound}.

\subsection{Entanglement bound from fractal dimensions}
\label{sec:entanglement_bound}
We have shown both analytically and numerically that at large $\lambda$, the entanglement of the RBM states is suppressed. We argued that the wavefunction amplitude generically becomes more non-uniform in the Ising basis, and this limits the entanglement of the RBM state at large $\lambda$. Now we concretize this argument by explicitly computing the ensemble-averaged fractal dimension, which bounds the entanglement entropy. The analysis will also develop understandings of the distinction between states in the near-maximal entropy regime and Haar-random states, as well as the deviation of level statistics from random matrix results in the entanglement spectrum of large $\lambda$.

The fractal dimension of order $q$ in the Ising basis is defined as follows~\cite{Evers:2008}:
\begin{equation}\label{eq: Dq_defn}
    D_q =\frac{1}{N\log 2}\frac{\log \text{IPR}_q}{1-q},\quad \text{IPR}_q=\frac{\sum_{\bs}|\Psi(\bs)|^{2q}}{(\sum_{\bs}|\Psi(\bs)|^2)^q},
\end{equation}
where $\Psi(\bs)$ is the wavefunction in the Ising basis: $\Psi(\bs)\equiv\braket{\bs}{\Psi}$ and IPR stands for the inverse participation ratio. %\TNc{1) Is the minus sign a typo? 2) Why average symbol in the definition? 3) In the definition, we should let $\bs$ be some general basis vector. 4) It may be nice to be more general and define $D_q$ with prefactor $1/(N\log d)$ where $d$ is the local Hilbert space dimension of the qdit dof.}
Ref.~\cite{multifractality2020} shows that $D_q$ bounds the $q$-th R\'{e}nyi entropy as (see Appendix \ref{app:proof_of_bound} for a proof):
\begin{equation}\label{eq: Dq_bound}
    \frac{S_{q}(A)}{N}\le D_q \log 2.
\end{equation}
Note that $D_q$ is basis-dependent, although the inequality holds for any basis in which subregime $A$ and its complement $B$ are not entangled. The inequality is saturated if the Schmidt basis is used to express the wavefunction components. In our later discussions, we will use the ensemble-averaged version of this bound: $\overline{S_q(A)}/N\le \overline{D_q}\log 2$.  %\TNc{Do we need to ensure that we're taking the average over only one variable and fixing the other for the inequality to hold? e.g. $\overline{S_q}\rvert_{D_q} \leq D_q N \log d$ or $\overline{D_q}\rvert_{S_q} \geq S_q/N \log d$. Ref seemed ambiguous, but also didn't read it in excrutiating detail.}

For our zero-bias RBM ensemble, it is convenient to use the Ising basis in which the wavefunction components are defined. We will analytically compute the large-$N$, ensemble-averaged value of $D_q$. For simplicity, we will only consider the limit of $(u,v)\rightarrow (0,\infty)$, which yields purely imaginary weights: $w_{mj}=i\tilde{w}_{mj}$ with $\tilde{w}_{mj}$ being Gaussian random variables of variance $v^2/N$. Once again, we assume small fluctuations in the numerator and denominator of IPR$_q$, and approximate the average of the $\log(\text{IPR}_q)$ by the $\log$ of the ratio of the averaged numerator and denominator. We begin with
\begin{equation}
\begin{split}
    \overline{\sum_{\bs}|\Psi(\bs)|^{2q}}&=\sum_\bs\prod_m\overline{\cos^{2q}(\widetilde{W}^m_{\bs})}\\
    &=\sum_{\bs} \left[ \frac{1}{2^{2q}} \sum_{k=-q}^{q}\binom{2q}{q+k}\overline{e^{i2 k \widetilde{W}_{\bs}^m}}\right]^M,
\end{split}
\end{equation}
where we have defined $\widetilde{W}^m_{\bs}=\sum_{\bs}\tilde{w}_{mj}s^j$, used the notation for binomial coefficient, and invoked independence of the terms in the product. In the large-$v$ limit, the sum is dominated by the $k=0$ term. Thus, we have the approximate result:
\begin{equation}
    \overline{\sum_{\bs}|\Psi(\bs)|^{2q}}\approx 2^N\left[\frac{1}{2^{2q}}\binom{2q}{q}\right]^M.
\end{equation}
Assuming both $\sum_{\bs}|\Psi(\bs)|^{2q}$ and the squared norm $\sum_\bs \abs{\Psi(\bs)}^2$ have small fluctuations, and using our previous result for the average squared norm, we can obtain:
\begin{equation}\label{eq: Dq_approx}
    \overline{D_q}\approx 1-\frac{q\lambda}{1-q}+\frac{1}{\log 2}\frac{\lambda}{1-q}\log \binom{2q}{q}.
\end{equation}
A remarkable observation is that the large-$N$ analytic calculation of $\overline{S_2(N/2)}$ for $\lambda > \lambda_c/2$ saturates the bound given by Eq.~\eqref{eq: Dq_bound} for $q = 2$, i.e., $\overline{S_2(N/2)}/(N\log 2) = \overline{D_2} =  1-\lambda\left(\frac{\log 3-\log 2}{\log 2}\right)\equiv 1-\lambda/\lambda_c$. This suggests that the underlying reason for the decrease of the entanglement entropy at large $\lambda$ is the wavefunction becomes more non-uniform with a smaller fractal dimension.

We now investigate the validity of our approximation for $\overline{D_q}$ by studying the fluctuations in $\sum_{\bs}|\Psi(\bs)|^{2q}$, which we took to be small in Eq.~\eqref{eq: Dq_approx}. We can compute $\overline{(\sum_{\bs}|\psi|^{2q})^2}$:
\begin{equation}
\begin{split}
    &\ \overline{(\sum_{\bs}|\psi|^{2q})^2}\\
    =&\sum_{\bs_1,\bs_2}\overline{\prod_m\cos^{2q}\left(\widetilde{W}^m_{\bs_1}\right)\cos^{2q}\left(\widetilde{W}^m_{\bs_2}\right)}\\
    =& \sum_{\bs_1,\bs_2} \prod_m \frac{1}{2^{4q}}\sum_{k,l=-q}^{q}\binom{2q}{q+k}\binom{2q}{q+l}\overline{e^{i2(k\widetilde{W}^m_1 + l\widetilde{W}^m_2})}\\[1em]
    \equiv& \sum_\phi e^{-N[\mathcal{E}(\phi)-\mathcal{S}(\phi)]},
\end{split}
\end{equation}
where once again, we recast the ensemble-averaged quantity in the form of a statistical mechanics partition function. As in previous calculations, the symmetries of the ansatz constrain the resulting free energy to depend only on $\phi\equiv (\bs_1\cdot \bs_2)/N$. Evaluating the average of the exponential term in the sum yields a term like $e^{-2v^2(k^2+l^2+2kl\phi)}$. In the large-$v$ limit, the $(k,l)$ configurations that dominate the sum depend on the value of $\phi$. There are two possible cases that must be considered separately: (1) $\abs{\phi}<1$ and (2) $\phi = \pm 1$. For $|\phi|< 1$, the dominating term is $k=l=0$. The free energy is:
%\TN{testing multiparagraph comment.
%does it work?}
\begin{equation}
    \mathcal{F}(\phi)=-2\lambda \log\left[\frac{1}{2^{2q}}\binom{2q}{q}\right]-\log 2-H_b\left(\frac{1-\phi}{2}\right),
\end{equation}
where $H_b$ is the binary entropy function in the natural log basis that we have introduced before. Note that for this case, the dependence on $\phi$ is contained entirely in the entropic term. This term yields a local minimum of the free energy at $\phi=0$:
\begin{equation}
    \mathcal{F}(\phi=0)=-2\lambda\log \left[\frac{1}{2^{2q}}\binom{2q}{q}\right]-2\log 2.
\end{equation}
In the other case for $\phi=\pm 1$, all terms with $k=\mp l$ can contribute in the large-$v$ limit. Using combinatorial identities, one can compute the energy density and find that the expression for both $\phi=\pm 1$ is
\begin{equation}
\begin{split}
    \mathcal{E}(\phi=\pm 1)&\approx-\lambda \log\sum_{k=-q}^q\frac{1}{2^{4q}}\binom{2q}{q+k}\binom{2q}{q-k}\\
    &=-\lambda\log \left[\frac{1}{2^{4q}}\binom{4q}{2q}\right],
\end{split}
\end{equation}
while the corresponding free energy is given by
\begin{equation}
    \mathcal{F}(\phi=\pm 1)\approx -\lambda\log \left[\frac{1}{2^{4q}}\binom{4q}{2q}\right]-\log 2.
\end{equation}
Given the expressions for the free energy in the two cases, we may now study the fluctuations in $\sum_{\bs}|\Psi(\bs)|^{2q}$ in different regimes. If the $\phi = 0$ minima dominates $\overline{(\sum_{\bs}|\Psi(\bs)|^{2q})^2}$, then the fluctuations of $\sum_{\bs}|\Psi(\bs)|^{2q}$ are small, and our calculation of $\overline{D_q}$ in Eq.~\eqref{eq: Dq_approx} is accurate to leading order in $1/N$. Namely, up to higher order terms in $1/N$,  we have $\log \overline{(\sum_{\bs}|\Psi(\bs)|^{2q})^2}\approx\log (\overline{\sum_{\bs}|\Psi(\bs)|^{2q}})^2$.
\begin{figure}[t]
    \centering
    \includegraphics[width=1\linewidth]{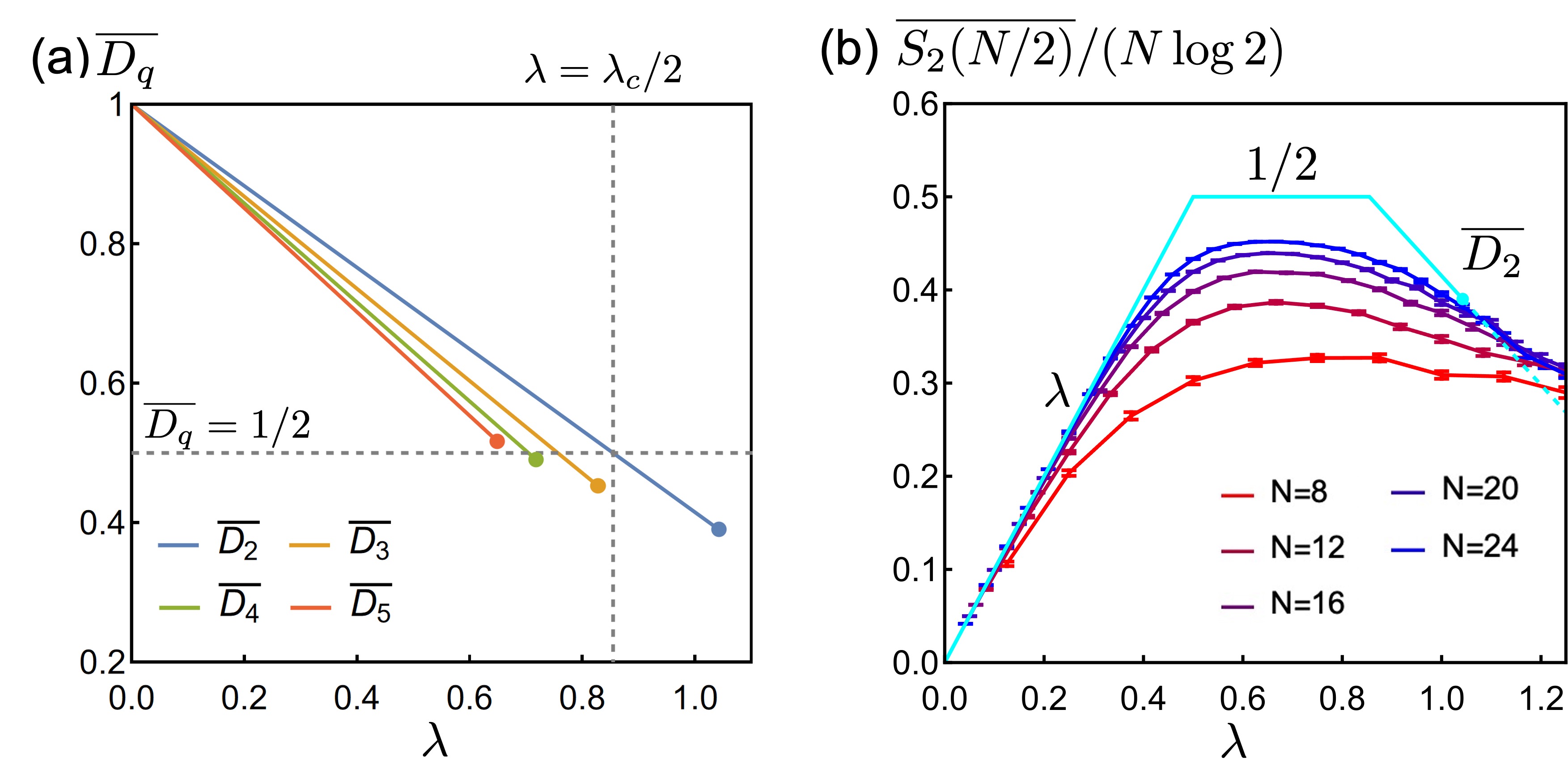}
    \caption{(a). Ensemble-averaged fractal dimensions $\overline{D_q}$ in the large-$N$ limit as a function of $\lambda$ for $(u,v)=(0,\infty)$. The solid dots are where the large-$N$ calculation becomes less accurate. (b). The average half-system second R\'enyi entropy for $(u,v)=(0,4)$ with different system size $N$ and $\lambda$. Each point is taken from the averaging over 100 random samples. The cyan line shows the bound of the average second R\'{e}nyi entropy for different $\lambda$, which is saturated in analytic large-$N$ calculations. Especially, the bound from the fractal dimension $\overline{D_2}$ can explain the reduction of the second R\'{e}nyi entropy at large $\lambda$.}
    \label{fig:fractal_dim}
\end{figure}
We can derive the condition on $\lambda$ such that the dominant free-energy minimum is at $\phi=0$ and our approximation for $\overline{D_q}$ is valid for:
\begin{equation}
    \lambda <\frac{\log 2}{\log [(4q)!]-4\log [(2q)!]+4\log (q!)}.
\end{equation}
Fig.~\ref{fig:fractal_dim}(a) shows $\overline{D_q}$ as a function of $\lambda$ up to the critical value when the fluctuation of $|\Psi(\bs)|^{2q}$ becomes large. The fractal dimension $\overline{D_q}$ for different $q$ provide an additional upper bound for the average $q$-th R\'{e}nyi entropy. For the second R\'{e}nyi entropy, we can see that this bound determines the sub-maximal behavior of entanglement for $\lambda>\lambda_c/2$ as shown in Fig.~\ref{fig:fractal_dim}(b). For $q=2$, we are able to compute the bound in the large-$N$ calculation up to $\lambda=\frac{\log 2}{\log 35/18}\approx 1.04$. Therefore, we can confirm the decreasing entanglement behavior for $\lambda>\lambda_c/2\approx 0.85$ since in this regime $\overline{D_q}<1/2$, meaning the half-system entropy is sub-maximal. This could also potentially explain the comparison of the finite-$N$ numerics to the analytic large-$N$ results at $\lambda=1.25$ [see Fig.~\ref{fig: numeric_Page_curves_s2}(c)]: for $\lambda>\frac{\log 2}{\log 35/18}$, the $\overline{D_2}$ bound that is saturated for the average second R\'{e}nyi entropy can take a different value from the current analytic calculations. In this regime, we expect the average second R\'{e}nyi entropy may also deviate from the large-$N$ analytic results. 

The fractal dimensions  provide an important measure for distinguishing states in the near-maximal regime from generic random states. For $1/2<\lambda<\lambda_c/2$ and $(u,v)=(0,\infty)$, we have computed that $\overline{D_q}<1$ for $q\ge 2$. Such states are usually called non-ergodic extended states in contrast with a localized state with $D_q=0$ and an ergodic state (such as a generic random state) with $D_q=1$. Hence, the fractal dimensions sharply distinguish these RBM states from a generic random state. 

Let us also comment on the the multifractal behavior; namely, the $q$-dependence of the bound for $q$-th R\'{e}nyi entropy. This suggests that the transition point to sub-maximal entanglement also has a $q$-dependence, and for the the calculation of second R\'{e}nyi entropy in the near-maximal entanglement regime,  the higher R\'{e}nyi entropies can indeed deviate from those of a generic random state. Although the \textit{spectrum} of the reduced density matrix in this regime is similar to that of a random quantum state for most of the eigenvalues, the higher R\'{e}nyi entropies are sensitive to deviations of the few large eigenvalues.

%\XLc{I think somewhere in this section we need some discussion about the Fig. 12d.}
Finally, we discuss the implications of the fractal dimension bound on the level statistics of the entanglement Hamiltonian in the large $\lambda$ case of Fig.~\ref{fig:levelstats}(d). As suggested by the large-$N$ calculation of $\overline{D_2}$, for $(u,v)=(0,\infty)$, the average half-system second R\'{e}nyi entropy saturates the $\overline{D_2}$ bound to leading order in large-$N$ limit. The saturation behavior suggests that for such a random RBM state the computational basis is close to the Schmidt basis (see Appendix \ref{app:proof_of_bound}). Heuristically, we can write the reduced density matrix of subregime $A$ of half-system size as
\begin{equation}
    \rho_A \simeq \frac{1}{\mathcal{N}}\left[\left( e^{-\sum_{ij}J^{ij}\sigma_i^z\sigma_j^z + \dots} \right) + \delta\rho\right],
\end{equation}
where $\mathcal{N}$ is a normalization factor. The first term contains all possible Pauli string operators from $\sigma^z_i$ such that this part is diagonal in the Ising basis. The second term of $\delta\rho$ contains corrections that are unimportant to lowest order in the large-$N$ calculations of $\overline{D_2}$ and $\overline{S_2}$. We can explain the part of entanglement spectrum corresponding to large eigenvalues of $\rho_A$ in Fig.~\ref{fig:levelstats}(d) with Poisson-like level statistics as coming from the part that is diagonal in the Ising basis. On the other hand, $\delta\rho$ produces the part of entanglement spectrum with GOE level statistics.

\subsection{Relation to quantum state designs}
The sharp distinction with Haar-random quantum states in the near-maximal entropy regime suggests that the RBM ensemble cannot form a quantum state design~\cite{Ambainis:2007}, which measures the similarity of an ensemble to the Haar-random ensemble. An ensemble of states represented by density matrix $\rho$ in a $d$-dimensional Hilbert space forms a quantum state $t$-design if $\overline{\rho^{\otimes t}}=\rho_{\text{Haar}}^{(t)}$, where $\rho_{\text{Haar}}^{(t)}=\int d\psi (|\psi\rangle\langle\psi|)^{\otimes t}$ and the integration is taken on the uniform Haar measure over all possible states in the $d$-dimensional Hilbert space. In this work, we focus on the thermodynamic limit of $N\rightarrow\infty$. We can define the ensemble to form a quantum state $t$-design in the large-$N$ limit if for any positive $\epsilon$, we can have a $N_0$ such that for $N>N_0$:
\begin{equation}
    \Big{\|} \overline{\rho^{\otimes t}}-\int d\psi (|\psi\rangle\langle\psi|)^{\otimes t} \Big{\|}_1<\epsilon,
    \label{eq:definition-of-design}
\end{equation}
where $\|...\|_1$ is the trace norm. Since the above trace norm decreases upon partial trace, from this definition, in the large-$N$ limit, if the ensemble forms a quantum state $t$-design, then it also forms $t'$-design for any $t'<t$. In the following, we will show that the zero-bias random RBM ensemble in the large-$N$ limit cannot form a quantum state $1$-design (and hence more generally cannot form a quantum state $t$-design).

A symmetry argument can be made to prove that any zero-bias random RBM ensemble cannot be a quantum state design. First, note that up to a normalization factor, the density matrix for a general zero-bias RBM state is as follows:
\begin{align}
    \ketbra{\Psi}{\Psi}= \sum_{\bs_1,\bs_2} \prod_{m}\cosh W_{\bs_1}^{m}\cosh(W_{\bs_2}^{m})^{*}\ketbra{\bs_1}{\bs_2}
\end{align}
The density matrix has matrix elements
\begin{equation}
    \rho_{\bs_1\bs_2}\sim \prod_{m}\cosh W_{\bs_1}^{m}\cosh(W_{\bs_2}^{m})^{*}
\end{equation}
The symmetry of the $\cosh()$ in the wave function implies symmetry of the matrix elements
\begin{equation}
    \rho_{\bs_1\bs_2}\equiv \rho_{\bs_1\overline{\bs_2}},
\end{equation}
where $\overline{\bs_2}\equiv\{-s_{2}^1,-s_{2}^2,...,-s_{2}^N\}$. There is a null space of this density matrix spanned by $|\boldsymbol{s}_{-}\rangle=\frac{1}{\sqrt{2}}(|\boldsymbol{s}\rangle-|\overline{\boldsymbol{s}}\rangle)$. The projectior onto this null space has sufficiently different expectation values in the zero-bias RBM ensemble and the Haar-random ensemble. Therefore, upon averaging, the density matrix cannot approach $\rho_{\text{Haar}}^{(1)}=\mathbb{1}/2^N$ since $\overline{\rho_{\bs\bs}}=\overline{\rho_{\bs\overline{\bs}}}$. Hence, a zero-bias RBM ensemble cannot form a 1-design. 

Furthermore, one can show that the RBMs do not form quantum state designs even in the subspace spanned by symmetric basis $\ket{\bs_{+}}=\frac{1}{\sqrt{2}}(\ket{\bs}+\ket{\overline{\bs}})$, which is invariant under $\ket{\bs}\rightarrow\ket{\overline{\bs}}$. This is ultimately because of the high statistical correlation between the wavefunction components for $\ket{\bs_{+}}$ and $\ket{\bs_{+}'}$. For example, assuming the norm fluctuations are small, we can consider the following element of the averaged density matrix:
\begin{equation}
    \overline{\rho_{\bs_{+}{\bs_{+}}'}}\sim \cosh^M\left((\bs \cdot \bs')(u^2+v^2)/N\right)
\end{equation}
In the following discussion, $\rho$ represents density matrix within the considered symmetric subspace. Since the density matrix will be normalized, each diagonal element shall be normalized to $\mathbb{1}/2^{N-1}$ (the dimension of the symmetric subspace is $2^{N-1}$). For $(\bs\cdot \bs')/N \approx 1$ (such as when $\bs$ and $\bs'$ differ by a small $\mathcal{O}(1)$ number of spins), the matrix elements remain of the same order as the diagonal element and the density matrix cannot approach $\rho_{\text{Haar}}^{(1)}=\mathbb{1}/2^{N-1}$. More precisely,  the expectation value of $\overline{\rho}-\rho_{\text{Haar}}^{(1)}$ can be positive and of the order of $1/2^{N-1}$ upon evaluating in the superposition state of $\frac{1}{\sqrt{2}}(|\bs_+\rangle+|\bs'_+\rangle)$ for $(\bs\cdot\bs')/N\approx 1$. There can be at least $ \sim 2^{N-2}$ such states that are orthogonal, and the projector onto the spanned subspace has sufficiently different expectation values in the zero-bias RBM ensemble and the Haar-random ensemble. Therefore, from the definition of Eq.~\eqref{eq:definition-of-design}, a zero-bias RBM ensemble does not form a 1-design in the symmetric subspace. It is worth mentioning that the above discussion is at large-$N$ limit for finite $(u,v)$ and should also apply to the near-maximal entanglement regime, where large-$N$ limit is taken before taking $(u,v)\rightarrow (0,\infty)$. %It is interesting that, with a polynomial number of classical parameters, we are able to efficiently generate a quantum-state ensemble that shares many properties of generic random states but still is not a quantum state design. 
%It is interesting that we are able to recover many properties of generic random states for an ensemble that is not a quantum state design. The ability to demonstrate a high-degree of entanglement entropy without mirroring the moments of the distribution of random quantum states is notable from a computational perspective. The nearly maximally-entangled RBMs states share many of the same bipartite entanglement characteristics with a random quantum state, but are classically simulable in polynomial time. In other words, our results show that bipartite entanglement, characterized by the von Neumann entropy and R\'enyi entropies, is \textit{not} the limiting factor in the efficient simulation of quantum states.

\section{Conclusion and outlook}
\label{sec:conclusions}

The curse of dimensionality is a profound conceptual and practical obstacle in theoretical many-body physics and other sciences. Machine learning has shown promise in circumventing this barrier by providing efficient (and ideally, unbiased) parameterizations of data sets in high-dimensional spaces. The RBM ansatz for many-body wavefunctions considered in this work is one example of such a parameterization.

In this work, we have used tools from statistical physics to methodically characterize a generic ensemble of random RBM wavefunctions. Our findings confirm that not only can RBMs efficiently encode volume-law entangled states, in agreement with previous studies, but they do so \textit{generically}, which sharply distinguishes them from other wavefunction ansatzes such as tensor network states. Importantly, we have demonstrated that random RBM states exhibit qualitatively different behavior as one tunes the parameters of the Gaussian random weights and the ratio of the number of hidden spins with the number of physical spins. Notably, in the thermodynamic limit $N\to\infty$, we find a parameter regime where the entanglement is maximal to lowest order in $1/N$. Interestingly, finite-size scaling analysis of the ensemble-averaged second R\`{e}nyi and von Neumann entropies shows that the deficit between the half-system entropy and the maximal value of $\left(N/2\right) \log 2$ is close to that of a Haar-random state. However, we find that even in this parameter regime, random RBM states are not equivalent to random quantum states as revealed by their fractal dimensions and the fact that they do not form a quantum state design.

It is intriguing that both Ref.~\cite{Deng:2017} and our results indicate that the averaged bipartite entanglement entropy for an RBM wavefunction ensemble is maximized at a finite value of $\lambda$. This stands in apparent contradiction with the intuition that RBMs with more hidden neurons have greater representational power, and can encode functions with more complex correlations. This appears to suggest an interesting limitation of the RBM ansatz: even if "wide" RBMs with more hidden neurons have greater representational power, wavefunctions that are highly entangled may comprise a fractionally smaller subspace of the set of \textit{all} wavefunctions that could be encoded by the RBM. This conjecture would explain the apparent mismatch between the high expressivity of wide RBMs and the non-monotonic relationship between the \textit{averaged} bipartite entanglement and $\lambda$. The practical implications of this would be that past a certain $\lambda$, adding more hidden neurons would not provide an advantage in targeting highly entangled wavefunctions during training. That being said, one prior investigation in a classical machine learning context showed that deeper neural network architectures can express more complex functions on average~\cite{raghu2017expressive}. It is possible that quantum states encoded by generic deeper RBMs can exhibit different behavior of entanglement as one increases number of hidden neurons. Developing a more concrete understanding of the typical training landscape for RBM wavefunctions at larger values of $\lambda$ is an interesting topic for future research.

Our results pave the way for understanding properties of generic neural network quantum states and provide insights
on future directions. First, our results reveal the richness of the RBM ansatz through the breadth of qualtitatively different quantum wavefunctions encoded by random RBMs. Our detailed characterization of a generic RBM ensemble provides guidance in the design and initialization of RBM quantum states for various tasks. Furthermore, our analytic approach may be straightforwardly generalized to explore the full RBM parameter space including the bias fields and to investigate other physical properties of interest (such as higher R\'enyi entropies). Such studies can facilitate a more detailed and fruitful understanding of the RBM quantum states in the future.  
\begin{acknowledgments}
We acknowledge useful discussions with Adam Nahum and Giuseppe De Tomasi. XQS acknowledges support from the Gordon and Betty Moore Foundation's EPiQS Initiative through Grant GBMF8691. MOF is supported by the Air Force Office of Scientific Research through grant No. FA9550-16-1-0334. XLQ is supported by the National Science Foundation under grant No. 2111998, and the Simons Fundation. This work is also supported in part by the DOE Office of Science, Office of High Energy Physics, the grant de-sc0019380. This work was partially finished when XLQ is visiting the Institute for Advanced Study, Tsinghua University (IASTU). XLQ would like to thank IASTU for hospitality. This work made use of the Illinois Campus Cluster, a computing resource that is operated by the Illinois Campus Cluster Program (ICCP) in conjunction with the National Center for Supercomputing Applications (NCSA) and which is supported by funds from the University of Illinois at Urbana-Champaign. The research also made use of the computational resources provided by the Stanford Research Computing Center. The contributions of MOF and XLQ were completed in part at the Aspen Center for Physics, which is supported by National Science Foundation grant PHY-1607611.
\end{acknowledgments}
\appendix

\begin{widetext}
\section{Convenient formulas for ensemble-averaging}\label{app: correlators}
In the main text, the calculation of the ensemble-averaged $\overline{\braket{\Psi}{\Psi}}$, $\overline{Z_0}$ and $\overline{Z_1}$ essentially boil down to the calculation of the following correlators:
\begin{equation}
    \overline{\cosh(W^{m}_{\bs_1})^{*} \cosh (W^{m}_{\bs_2})}\qquad\text{and}\qquad \overline{\cosh(W_{\bs_1}^{m})^{*}\cosh(W_{\bs_2}^{m})^{*}\cosh(W_{\bs_3}^{m})\cosh(W_{\bs_4}^{m})},
\label{eq: correlators}
\end{equation}
The above correlations will ultimately depend on the correlators between random variables $(W_{\bs_{i}}^{m})^{*}$ and $W_{\bs_{j}}^{m}$. Recall that $W^m_\bs \equiv \sum_j w_{m j} s^j$ and $w_{m j} = w^{R}_{m j} + i w^{I}_{m j}$ for $w^{R}_{m j},  w^{I}_{m j}$ random real Gaussian variables drawn from $\SN(0,\sigma_R^2)$ and $\SN(0,\sigma_I^2)$, respectively. Therefore we have $\overline{w_{m j}^{R} w_{m k}^{R}}= \sigma_R^2 \delta_{jk}$, $\overline{w_{m j}^{I} w_{m k}^{I}} = \sigma_I^2 \delta_{jk}$, and $\overline{w_{m j}^{R} w_{m k}^{I}} = 0$. From these properties, it is straightforward to obtain:
\begin{equation}
\begin{split}
    &\overline{W^m_{\bs_i}\left(W^m_{\bs_j}\right)^* } =(\bs_i\cdot \bs_j) (u^2 + v^2) /N,\\
    &\overline{ W^m_{\bs_i} W^m_{\bs_j} } = 
    \overline{ \left(W^m_{\bs_i}\right)^* \left(W^m_{\bs_j}\right)^* } = (\bs_i\cdot \bs_j) (u^2 - v^2) /N,
\end{split}
\label{eqs: Wcorrelators}
\end{equation}
where we have defined $u^2\equiv \sigma_R^2/N$ and $v^2\equiv \sigma_I^2/N$.

The strategy for computing the correlators in \eqref{eq: correlators} is as follows: 1) expand the $\cosh()$ terms using exponential functions 2) recast the product of $\cosh()$ terms in the correlator as a partition function summing over classical binary variable configurations (which will represent the possible combinations of exponentials with positive and negative arguments), and 3) compute the ensemble average directly by taking advantage of the Gaussianity of the random weights.

We will walk through the procedure in detail for $\overline{\cosh(W^{m}_{\bs_1})^{*} \cosh (W^{m}_{\bs_2})}$, and then report the results from straightforward generalization to $\overline{\cosh(W_{\bs_1}^{m})^{*}\cosh(W_{\bs_2}^{m})^{*}\cosh(W_{\bs_3}^{m})\cosh(W_{\bs_4}^{m})}$. First, we perform the expansion over exponentials,
\begin{equation}
    \overline{\cosh(W^{m}_{\bs_1})^{*} \cosh (W^{m}_{\bs_2})} = \frac{1}{4}\overline{\Big(\exp\left(W^{m}_{\bs_1}\right)^* + \exp\left(-W^{m}_{\bs_1}\right)^*\Big)  \Big( \exp\left(W^{m}_{\bs_2}\right)^* + \exp\left(-W^{m}_{\bs_2}\right)^*\Big)}.
\end{equation}
Next, we introduce auxiliary binary variables $\rho_i$ which take on values in $\{-1,+1\}$ to succinctly rewrite the above expression:
\begin{align}
      \overline{\cosh(W^{m}_{\bs_1})^{*} \cosh (W^{m}_{\bs_2})}&=\frac{1}{4}\sum_{\{\rho_i=\pm 1\}}\overline{\exp\Big(\rho_1(W^{m}_{\bs_1})^{*}+\rho_2 W^{m}_{\bs_2} \Big)}.
\end{align}
Now we will take advantage of the following useful result for independent Gaussian random variables $x_k$:
\begin{align}\label{eq: Gaussian_property}
    \overline{\exp\left( \sum_k c_k x_k\right)} = \exp\left[\frac{1}{2}\overline{\left(\sum_k c_k x_k\right)^2} \right].
\end{align}
Since $W^{m}_{\bs_i}$ is a linear combination of independent random Gaussian variables (and is itself a Gaussian random variable), the above property may be invoked. These ensemble average can accordingly be brought inside the argument of the exponential and computed using the correlators in Eq.~\eqref{eqs: Wcorrelators}:
\begin{align}
    \begin{split}
        \overline{\cosh(W^{m}_{\bs_1})^{*} \cosh (W^{m}_{\bs_2})}&= \frac{1}{4}\sum_{\{\rho_i=\pm 1\}}\exp\left[\frac{1}{2}\overline{\Big(\rho_1(W^{m}_{\bs_1})^{*}+\rho_2 W^{m}_{\bs_2} \Big)^2}\right]\\
        &= \frac{1}{4}\sum_{\{\rho_i=\pm 1\}}\exp\Big(f(\{\rho_i\},\bs_1,\bs_2)\Big)
        \label{eq: 2cosh_correlator},
    \end{split}
\end{align}
where the function $f$ is defined to be
\begin{equation}
\begin{split}
    f(\{\rho_i\},\bs_1,\bs_2)&\equiv\frac{1}{2}\overline{\Big(\rho_1(W^{m}_{\bs_1})^{*}+\rho_2 W^{m}_{\bs_2} \Big)^2}\\
    &=(u^2-v^2)+\rho_1\rho_2(\bs_1\cdot\bs_2)(u^2+v^2)/N,
    \label{eq: f_2cosh_correlator}
\end{split}
\end{equation}
Performing the analogous calculation for the other correlator in Eq. \eqref{eq: correlators}, one has
\begin{equation}
\begin{split}
    \overline{\cosh(W_{\bs_1}^{m})^{*}\cosh(W_{\bs_2}^{m})^{*}\cosh(W_{\bs_3}^{m})\cosh(W_{\bs_4}^{m})}&=\frac{1}{16}\sum_{\{\tau_i=\pm 1\}}\exp[g(\{\tau_i\},\bs_1,\bs_2,\bs_3,\bs_4)\})],
\end{split}
\label{eq: 4cosh_correlator}
\end{equation}
where
\begin{equation}
\begin{split}
 g(\{\tau_i\},\bs_1,\bs_2,\bs_3,\bs_4)&\equiv \frac{1}{2}\overline{[\tau_{1}(W_{\bs_{1}}^{m})^{*}+\tau_{2}(W_{\bs_{2}}^{m})^{*}+\tau_{3}W_{\bs_{3}}^{m}+\tau_{4}W_{\bs_{4}}^{m}]^2}\\
 &=2(u^2-v^2)+\sum_{k\in\{1,2\};l\in\{3,4\}}\tau_{k}\tau_l (u^2+v^2)(\bs_k\cdot \bs_l)/N\\
 &+\tau_1\tau_2 (u^2-v^2)(\bs_1\cdot \bs_2)/N+\tau_3\tau_4 (u^2-v^2)(\bs_3\cdot \bs_4)/N.
\end{split}
\label{eq: g_4cosh_correlator}
\end{equation}
\section{Derivation of $\overline{\abs{\Psi}^2}$ [Eq.~\eqref{eq: logAvgNormSq_red}]}
\label{app:norm_square_av}
Using the correlator in Eq.~\eqref{eq: 2cosh_correlator}, we can write the average norm of the Gaussian RBM ensemble as
\begin{align*}
    \overline{\braket{\Psi}{\Psi}} &= \sum_{\bs} \prod_{m=1}^M \overline{\cosh(W_\bs^m)^* \cosh(W_\bs^m)} \nonumber \\
    &= \sum_{\bs} \prod_{m=1}^M \frac{1}{4} \sum_{\{\rho_i=\pm 1\}} \exp\left[(u^2 - v^2) + \rho_1\rho_2(u^2 + v^2)\right].
\end{align*}
Explicitly summing over the $\{\rho_i\}$ variables, one has
\begin{align}
    \overline{\braket{\Psi}{\Psi}} &= \sum_{\{s\}} \prod_{m=1}^M \frac{1}{2}\left[\exp\left(2u^2\right) + \exp\left(2v^2\right) \right] \nonumber \\
     &= \frac{2^N}{2^M}\left[\exp(2u^2)+\exp(-2v^2)\right]^M.
\end{align}

\section{Derivation of $\overline{Z_0} =\overline{\braket{\Psi}{\Psi}^2}$ [Eq.~\eqref{eq: Z0_Gaussian} and Eq.~\eqref{eq: Z0_free_energy}]}
\label{app: Z0}
Here, we compute $\overline{Z_0} = \overline{\braket{\Psi}{\Psi}^2}$ and map it to a partition function of a coupled spin chain problem. Using the correlator in Eq.~\ref{eq: 4cosh_correlator}, we can write $\overline{Z_0}$ as
\begin{align*}
\begin{split}
    \overline{Z_0}&= \sum_{\bs_1,\bs_2}
    \prod_{m = 1}^M \overline{\cosh\left(W^m_ {\bvec{s_{1}}}\right)^*\cosh\left(W^m_{\bvec{\bs_1}}\right) \cosh\left(W^m_{\bvec{s_{2}}}\right)^*\cosh\left(W^m_{\bvec{\bs_2}}\right)}\\
    &=\frac{1}{16}\sum_{\{\tau_i=\pm 1\}}\exp[g(\{\tau_i\},\bs_1,\bs_2,\bs_1,\bs_2)\})].
\end{split}
\end{align*}
Explicitly summing over the $\{\tau_i\}$ variables gives Eq.~\eqref{eq: Z0_Gaussian}:
\begin{equation}
\begin{split}
    \overline{Z}_0=e^{2M(u^2-v^2)}\sum_{\bs_1,\bs_2}\Big{[}\frac{1}{4}(2+e^{2(u^2+v^2)}\cosh 4\phi u^2+e^{-2(u^2+v^2)}\cosh 4\phi v^2) \Big{]}^M,
\end{split}
\label{eq: Z0_result_1}
\end{equation}
where $\phi\equiv (\bs_1\cdot\bs_2)/N$. Since the terms in the sum really only depend on $\phi$, $\overline{Z_0}$ can be recast as a partition function whose free energy depends on a single effective degree of freedom $\phi$, with some degeneracy factor that represents the number of $(\bvec{\bs_1},\bvec{\bs_2})$ configurations that map to the same $\phi$, i.e.,
\begin{equation}
    \overline{Z_0}=\sum_{\bs_1,\bs_2}e^{-N\mathcal{E}(\phi)}=\sum_{\phi}\mathcal{D}(\phi)e^{-N\mathcal{E}(\phi)},
\end{equation}
where $\mathcal{D}(\phi)$ is the degeneracy factor. Comparing to Eq.~\eqref{eq: Z0_result_1} we can obtain the expression for $\mathcal{E}(\phi)$ in Eq.~\eqref{eq: Z0_energy}.

Now let us compute this degeneracy factor $\mathcal{D}(\phi)$. The dot product $N\phi = \bs_1\cdot \bs_2 = \sum_i s^i_1 s^i_2 $ counts the difference between the number of indices for which $s^i_1$ and $s^i_2$ agree and the number for which they disagree, and is equivalent to addition modulo 2 for two binary vectors. The number of indices that agree is $N(1+\phi)/2$ and the number of indices that disagree is $N(1-\phi)/2$. Hence, the degeneracy factor is
\begin{align}
  \mathcal{D}(\phi)=2^N\binom{N}{\frac{\phi+N}{2}} = 2^N\frac{N!}{\frac{N(1 + \phi)}{2}!\frac{N(1 -\phi)}{2}!}.
\end{align}
Now, $\overline{Z}_0$ may be rewritten as a partition function of a statistical mechanics problem that depends on the single parameter $\phi$:
\begin{equation}
    \overline{Z_0}=\sum_{\phi} \exp[-N \mathcal{F}(\phi)],
    \label{eq: Z0_free_energy_app}
\end{equation}
where the free energy density $\mathcal{F}(\phi)=\mathcal{E}(\phi)-\mathcal{S}(\phi)$ and the entropy density $\mathcal{S}(\phi)\equiv\frac{1}{N}\log \mathcal{D}(\phi)$ is given by
\begin{equation}
    \mathcal{S}(\phi)=-\frac{1-\phi}{2}\log \frac{1-\phi}{2}-\frac{1+\phi}{2}\log \frac{1+\phi}{2}+\log 2.
\end{equation}
where we have used Stirling's approximation.

\section{Derivation of $\overline{Z_1} \equiv \tr \left(X_A\overline{ \ketbra{\Psi}{\Psi}\otimes \ketbra{\Psi}{\Psi}}\right) $ [Eq.~\eqref{eq: spin_sum}]}\label{app: Z1_computation}
The general correlator in Eq.~\eqref{eq: 4cosh_correlator} may be used to compute $\overline{Z_1}$, which is given by Eq.~\eqref{eq: Z1_tensor}:
\begin{equation}
    \overline{Z_1}=\sum_{\bs_1,\bs_2}\left(\frac{1}{16}\sum_{\{\tau_i\}}\exp\left[g(\{\tau_i\},\bs_1,\bs_2,\bs_3,\bs_4)\right]\right)^{M}\delta_{\bs_2^A\bs_3^A}\delta_{\bs_1^B\bs_3^B}\delta_{\bs_2^B\bs_4^B},
\end{equation}
The delta tensors implement the following identifications (schematically depicted in Fig.~\ref{fig: Z1_tensor}):
\begin{equation}
\begin{split}
    \bs_1\cdot \bs_3&=\bs_{2}\cdot \bs_{4}=N\phi_A+|B|,\\
    \bs_{1}\cdot \bs_{4}&=\bs_{2}\cdot \bs_{3}=N\phi_B+|A|,\\
    \bs_1\cdot \bs_2&=\bs_3\cdot \bs_4=N(\phi_A+\phi_B),
\end{split}
\end{equation}
where we have used the same notation of $\phi_A$ and $\phi_B$ as in the main text.
Substituting these equations to the expression for $g(\{\tau_i\},\bs_1,\bs_2,\bs_3,\bs_4)$ in Eq.~\eqref{eq: g_4cosh_correlator} and explicitly summing over $\{\tau_i\}$ yields
\begin{equation}
    \overline{Z_1}=\sum_{\bs_1,\bs_2}e^{-N\mathcal{E}(\phi_A,\phi_B)}=\sum_{\phi_A,\phi_B}\mathcal{D}(\phi_A,\phi_B)e^{-N\mathcal{E}(\phi_A,\phi_B)},
\end{equation}
where $\mathcal{E}(\phi_A,\phi_B)$ may be computed to take the form in Eq.~\eqref{eq: energy_AB}. Analogous to the reasoning in the previous section, the degeneracy factor eventually yields the entropy density $\mathcal{S}(\phi_A,\phi_B)\equiv \frac{1}{N}\log \mathcal{D}(\phi_A,\phi_B)$, whose formula is presented in Eq.~\eqref{eq: entropy_AB}.
%\TN{
%\section{Fluctuations in $\text{IPR}_q$}
%When I did the computation by hand, as an intermediate step I got
%\begin{align}
%    \overline{\left(\sum_{\bs} \abs{\psi}^{2q}\right)^2} &=  \frac{1}{2^{4qM}} \sum_{\bs_1,\bs_2} \left( \sum_{kr} \binom{2q}{k}\binom{2q}{r} \exp\left(-2v^2 f(k,r,\phi) \right) \right)^M\\[1em]
%    f(k,r,\phi) &= 2q^2(1+\phi) - 2q(k+r)(1+\phi) + (k^2+r^2+2kr\phi)
%\end{align}
%so the dominant terms in the sum $\sum_{kr}$ would be those that minimize $f$
%}
\section{Proof of Eq.~\eqref{eq: Dq_bound}}\label{app:proof_of_bound}
In this section we shall present a proof of Eq.~\eqref{eq: Dq_bound}. We will use the normalized wavefunction here, namely $\tilde{\Psi}(\bs)=\Psi(\bs)/\sqrt{\sum_{\bs'}|\Psi(\bs')|^2}$. For a bipartition of the system as subregion $A$ and its compliment $B$, we can write the normalized state as a Schmidt decomposition:
\begin{equation}
    |\tilde{\Psi}\rangle=\sum_n\lambda_n^{1/2}|E^n_A\rangle\otimes|E^n_B\rangle,
\end{equation}
where $|E^n_{A/B}\rangle$ form an orthonormal basis of the Hilbert space in regime $A/B$, respectively. Using the definition of $D_q$ and $S_q$, the inequality of Eq.~\eqref{eq: Dq_bound} can be derived from the following inequality:
\begin{equation}
    \sum_{i_A,i_B}|\tilde{\Psi}_{i_A,i_B}|^{2q}\le \sum_{n}\lambda_n^q,
\label{eq:general_basis}
\end{equation}
where $\tilde{\Psi}_{i_A,i_B}\equiv(\langle i_A|\otimes\langle i_B|)|\tilde{\Psi}\rangle=\sum_n\lambda_n^{1/2}\langle i_A|E_A^n\rangle\langle i_B|E_B^n\rangle$ is the normalized wavefunction in given (orthonormal) computational basis $|i_A\rangle\otimes|i_B\rangle$. Here $|i_{A/B}\rangle$ is an orthonormal basis for the Hilbert space in subregion $A/B$. While in the main text the Ising basis is chosen, here we will keep the discussion general for any basis that is unentangled between subregion $A$ and $B$ and explore the condition for the inequality to saturate.

The proof of Eq.~\eqref{eq:general_basis} is as follows: First, we notice that
\begin{equation}
    \sum_{i_A,i_B}|\tilde{\Psi}_{i_A,i_B}|^{2q}\le\sum_{i_A}(\sum_{i_B}|\tilde{\Psi}_{i_A,i_B}|^2)^q,
\end{equation}
which is true since all $|\tilde{\Psi}_{i_A,i_B}|^2$ are non-negative numbers and the right hand has more non-negative terms when expanding to a polynomial of $|\tilde{\Psi}_{i_A,i_B}|^2$. This inequality saturates when for fixed $i_A$, there is only one nonzero $\tilde{\Psi}_{i_A,i_B}$. Notice that we can recast $\sum_{i_B}|\tilde{\Psi}_{i_A,i_B}|^2$ as
\begin{equation}
\begin{split}
    \sum_{i_B}|\tilde{\Psi}_{i_A,i_B}|^2&=\sum_{n,n',i_B}\lambda_n^{1/2}\lambda_{n'}^{1/2}\langle i_A|E_A^n\rangle\langle i_B|E_B^n\rangle\langle E^{n'}_B|i_B\rangle\langle E_A^{n'}|i_A\rangle\\
    &=\sum_n\lambda_n|\langle i_A|E_A^n\rangle|^2,
\end{split}
\end{equation}
where we have used the completeness of $|i_B\rangle$ and $\langle E_B^{n'}|E_B^n\rangle=\delta_{nn'}$. Now we can finish the proof of Eq.~\eqref{eq:general_basis}:
\begin{equation}
 \sum_{i_A,i_B}|\tilde{\Psi}_{i_A,i_B}|^{2q}\le\sum_{i_A}(\sum_{i_B}|\tilde{\Psi}_{i_A,i_B}|^2)^q=\sum_{i_A}(\sum_n \lambda_n|\langle i_A|E_A^n\rangle|^2)^q\le\sum_{n,i_A}\lambda_n^q|\langle i_A|E_A^n\rangle|^2=\sum_n\lambda_n^q,
\end{equation}
where we have used the following inequality for the probability distribution $p_n=|\langle i_A|E_A^n\rangle|^2$:
\begin{equation}
    \sum_n\lambda_n p_n\le \sum_n (\lambda_n^{q} p_n)^{1/q}.
\end{equation}
The inequality is saturated when only one of $p_n$ is nonzero. This means that $\{|i_A\rangle\}$ is the same as the Schmidt basis $\{|E_A^n\rangle\}$. From a similar approach (swapping subregion labels $A\leftrightarrow B$), we can also show that when the inequality of Eq.~\eqref{eq:general_basis} saturated, $\{|i_B\rangle\}$ is the same as the Schmidt basis $\{|E_B^n\rangle\}$. 

%\XQ{\section{Level statistics at generic $(u,v,\lambda)$}\label{app:generic_level}
%In the main text, we report calculations for $(u,v)=(0,4)$ where the wavefunctions are real and find level repulsion close to the GOE result for different $\lambda$. Here we will show that the behavior can be different from GOE when both $u$ and $v$ are nonzero. In this case, the RBM ensemble contain wavefunctions of complex value. As can be seen from the level }
%\begin{figure}
%    \centering
%    \includegraphics[width=0.45\textwidth]{level_stat_add.jpg}
%    \caption{Caption}
%    \label{fig:my_label}
%\end{figure}
\end{widetext}

\bibliography{bib}{}

\begin{thebibliography}{52}
\bibitem [{Notei()}]{Notei}%
  \BibitemOpen
  \bibinfo {note} {To properly minimize the free energy and find the global
  minimum, one should properly take into account possible local minima near the
  $\phi _A=1/2$ or $\phi _B=1/2$ line and the corner $\phi _A=\phi
  _B=1/2$.}\BibitemShut {Stop}%
\end{thebibliography}
\end{document}